\documentclass[11pt]{article}
\usepackage{amsfonts}
\usepackage{amsmath}
\usepackage{mathtools}
\usepackage{color}
\usepackage{multibib}
\usepackage{bm}
\usepackage{xurl}
\usepackage{setspace}
\usepackage{algorithm}
\usepackage[nohead]{geometry}
\usepackage{algcompatible}
\usepackage[round]{natbib}
\usepackage{tikz}  
\usepackage{calc}
\usepackage{hyperref}
\usepackage{multirow}
\usetikzlibrary{shapes.geometric, arrows, positioning,shapes,calc}
\usepackage{longtable}
\usepackage{arydshln}

\def\argmin{\mathop{\mbox{argmin}}}
\DeclareMathOperator*{\minimize}{\text{minimize}}

\def\A{{\mathbf{A}}}
\def\B{{\mathbf{B}}}
\def\D{{\mathbf{D}}}
\def\E{{\mathbf{E}}}
\def\M{{\mathbf{M}}}
\def\bM{{\mathbf{M}}}

\def\X{{\mathbf{X}}}

\def\U{{\mathbf{U}}}
\def\V{{\mathbf{V}}}

\def\I{{\mathbf{I}}}

\def\P{{\mathbf{P}}}

\def\R{{\mathbf{R}}}
\def\Z{{\mathbf{Z}}}
\def\bl{{\bm{\lambda}}}
\def\rank{\text{rank}}

\makeatletter
  \renewcommand*\env@matrix[1][*\c@MaxMatrixCols c]{%
    \hskip -\arraycolsep
    \let\@ifnextchar\new@ifnextchar
  \array{#1}}
\makeatother

\newtheorem{my_def}{Definition}
\newtheorem{ex}{Example}

\newtheorem{example}{Example}

\newpage

 
\begin{document}
\title{\ Hierarchical nuclear norm penalization for multi-view data}
\author{\large Sangyoon Yi\thanks{syi@stat.tamu.edu}, Raymond K.W. Wong\thanks{raywong@stat.tamu.edu}, Irina Gaynanova\thanks{irinag@stat.tamu.edu}
\medskip\\
{\large Department of Statistics, Texas A\&M University, College Station, TX 77843, USA} }
\date{}

\maketitle
\sloppy%

\textbf{Abstract:} The prevalence of data collected on the same set of samples from multiple sources (i.e., multi-view data) has prompted significant development of data integration methods based on low-rank matrix factorizations. These methods decompose signal matrices from each view into the sum of shared and individual structures, which are further used for dimension reduction, exploratory analyses, and quantifying associations across views. However, existing methods have limitations in modeling partially-shared structures due to either too restrictive models, or restrictive identifiability conditions. To address these challenges, we formulate a new model for partially-shared signals based on grouping the views into so-called hierarchical levels. The proposed hierarchy leads us to introduce a new penalty, hierarchical nuclear norm (HNN), for signal estimation. In contrast to existing methods, HNN penalization avoids scores and loadings factorization of the signals and leads to a convex optimization problem, which we solve using a dual forward-backward algorithm. We propose a simple refitting procedure to adjust the penalization bias and develop an adapted version of bi-cross-validation for selecting tuning parameters. Extensive simulation studies and analysis of the genotype-tissue expression data demonstrate the advantages of our method over existing alternatives. 

\strut \textbf{Keywords:} Bi-cross-validation; Data fusion; Low-rank matrix; Multi-source data; Optimization; Rank selection.

\section{Introduction}\label{sec:intro}
Technological advances in biomedical fields led to prevalence of multi-source data collected from the same set of samples, often called multi-view data. Our motivating example is the Genotype-Tissue Expression (GTEx) project \citep{gtex2020gtex} that collects gene expression data on the same individuals across multiple tissues (views). The gene-expression is often tissue-specific. For instance, the p53 gene is critical for cell regulation playing prominent role in cancer development \citep{tanikawa2017transcriptional}, however p53 tissue-specific expression makes it difficult to develop targeted therapies \citep{vlatkovic2011tissue}. It is thus crucial to identify patterns in gene expression that are shared and unique across multiple tissues. Traditional methods for signal extraction tend to be applied either separately to each view or to all views combined, thus failing to separate shared and unique parts of the signal.

Specifically, let $\X_{d}\in\mathbb{R}^{n \times p_{d}}$ be the observed data matrix for the $d$-th view with $n$ matched samples and $p_{d}$ measurements for $d=1,\dots,D$. We consider an additive error model
\begin{equation}\label{eq:ourmod}
\X_{d} = \M_{d}  + \E_{d}, \quad d=1,\dots,D,  
\end{equation}
where $\M_{d}$ is the signal matrix and $\E_{d}$ is the noise matrix. Successful extraction of the signals in the model~\eqref{eq:ourmod} requires additional structural assumptions. For single-view case, $\M_{d}$ is commonly assumed to be low-rank \citep{srebro2003weighted, candes2009exact, candes2013unbiased}. A direct application of this approach for multi-view data leads to low-rank assumption on either (i) each $\M_d$ or (ii) the concatenated signal $[\M_1\dots \M_D]$. The main limitation of (i) is that it not only ignores any structural relationships between the signals due to the matched samples, but also separately estimates signals without borrowing strength across views. The main limitation of (ii) is that it assumes a joint structure across all $\M_d$, resulting in all estimated $\widehat \M_d$ having the same rank. Therefore, it does not take into account possible heterogeneity across the views. In multi-view context, it is of interest to explore structural assumptions on $\M_d$ that allow both joint (thus accounting for matched samples) and individual (thus permitting differences across views) parts of the signal matrices. 

The simultaneous exploration of the joint and individual signals has been a prominent line of research for multi-view data \citep{lock2013joint,zhou2016group,yang2016non,feng2018angle,gaynanova2019structural}. The joint structure is a shared pattern across all views and defined as the intersection of the column spaces of $\M_{d}$ by JIVE \citep{lock2013joint}, COBE \citep{zhou2016group} and AJIVE \citep{feng2018angle}, while the individual structure is unique to a particular view adjusted after the joint structure, see Example \ref{ex:d2} in Appendix \ref{supp:method} for illustration. Despite these advances, many existing approaches \citep{lock2013joint,zhou2016group,yang2016non,feng2018angle} lack explicit definition of partially-shared structures when $D>2$, e.g., a signal structure shared by muscle and blood tissues, but not the skin tissue in the motivating GTEx example. This implies that the corresponding methods cannot take advantage of partially-shared structures for signal estimation, leading to rank mis-identification and worse signal estimation performance compared with methods that account for such structures \citep{gaynanova2019structural}. 

While several methods for identification of partially-shared structures have been proposed, they have limitations from modelling perspective. The penalized matrix factorization approaches by \cite{jia2010factorized} and \cite{van2011flexible} lack explicit formulation of underlying models, making it difficult to assess identifiability and provide interpretation. SLIDE \citep{gaynanova2019structural} requires orthogonality across individual and partially-shared signals for model identifiability. This implies that the SLIDE model is not always unique, causing issues in estimation as well as interpretation.

In addition to modeling issues (lack of partially-shared structures or lack of identifiability), existing methods \citep{jia2010factorized, van2011flexible, lock2013joint, zhou2016group,yang2016non, gaynanova2019structural} estimate the signals $\M_d$ based on explicit score-and-loading factorizations, leading to non-convex optimization problems. The convergence to the global optimum is not guaranteed and the obtained solution depends on the starting point. Thus, in practice, the obtained solution may be sub-optimal. The recently proposed BIDIFAC \citep{park2020integrative} and BIDIFAC+ \citep{lock2022bidimensional} are exceptions, as those methods provide a convex formulation via the use of nuclear norm penalization. Both are designed for bidimensionally-matched data, which includes multi-view data as a special case. However, BIDIFAC does not consider partially-shared structures. Furthermore, both BIDIFAC and BIDIFAC+ suffer from bias associated with nuclear norm penalization \citep{chen2013reduced,josse2016adaptive}, affecting their signal estimation performance. 
 
In this work, we address the limitations of existing models by both formulating an explicit model for partially-shared structures and avoiding identifiability issues. We also address the limitations of associated estimation approaches by avoiding scores and loadings factorization, and correcting for possible estimation bias. We achieve these goals as follows. First, we propose a recursive definition of partially-shared structures based on the new concept of hierarchical level. For $D=3$ views, the hierarchical levels are: (i) all three views (Level 1); (ii) any two views (Level 2); (iii) individual views (Level 3). The key idea is to work with the column spaces of $\bM_d$, and recursively consider column spaces at each level after accounting for previous levels. Our formulation includes JIVE, COBE and AJIVE models as a special case and avoids identifiability issues of SLIDE. Secondly, guided by our hierarchical levels, we propose a new penalty, hierarchical nuclear norm (HNN), for direct estimation of $\bM_d$ without scores and loadings factorization. To our knowledge, HNN is novel in the literature, and its hierarchical construction allow to restrict the ranks of the signals at each level. By varying the corresponding tuning parameters, the solution path encompasses a wide range of models with different patterns of shared signals. HNN is convex, but does not have a closed-form proximal operator. We address this challenge by adapting the dual block-coordinate forward-backward algorithm \citep{abboud2017dual}.  Third, we address penalization bias by a new type of refitting. As the proposed definition of partially-shared structures relies on the column spaces, we use HNN to estimate the column spaces, and then refit $\bM_d$ under the column space constraint, thus preserving the estimated shared and individual patterns. Finally, we take advantage of SURE criterion \citep{candes2009exact} to reduce the number of tuning parameters, and use the adapted version of the bi-cross validation \citep{owen2009bi} for model selection. In numerical studies, the resulting HNN method outperforms the competitors in both signal estimation and structure identification. 

\emph{Notation}. For a matrix $\A = (a_{ij})\in\mathbb{R}^{n\times p}$, let $\|\A\|_{F}=\sqrt{\sum_{i=1}^{n}\sum_{j=1}^{p} a_{ij}^{2}}$ be the Frobenius norm,  and $\|\A\|_{*} = \sum_{i=1}^{\text{min}(n,p)} \sigma_{i}(\A)$ be the nuclear norm, where $\sigma_{i}(\A)$ is the $i$-th largest singular value of $\A$. For the column space $\mathcal{C}(\A)$, let $\P_{\mathcal{C}(\A)}$ and $\P_{\mathcal{C}(\A)}^\perp$ be the projection matrices onto $\mathcal{C}(\A)$ and its orthogonal complement $\mathcal{C}(\A)^\perp$, respectively. Let $\mathbf{0}$ and $\I$ be the zero and identity matrix, respectively, with dimensions inferred from the context. Let $[\A_{1}\ \A_{2}]\in\mathbb{R}^{n \times (p_{1}+p_{2})}$ be formed by concatenating $\A_{1}\in\mathbb{R}^{n\times p_{1}}$ and $\A_{2}\in\mathbb{R}^{n\times p_{2}}$ column-wisely. Given the column spaces $\mathcal{C}_{1},\mathcal{C}_{2}\subset\mathbb{R}^{n}$ of matrices $\A_{1}$ and $\A_{2}$, define $S_{\mathcal{C}_{1}}^\perp(\mathcal{C}_{2})$ to be the column space of matrix $\P_{\mathcal{C}_{1}}^\perp\A_{2}$, that is $S_{\mathcal{C}_{1}}^\perp(\mathcal{C}_{2})=\mathcal{C}(\P_{\mathcal{C}_{1}}^\perp\A_{2})$. Let $\mathcal{C}_{1}+\dots+\mathcal{C}_{k}=\text{span}\{\mathcal{C}_{1},\dots,\mathcal{C}_{k}\}$ be subspace spanned by the union of sets of basis vectors of subspaces $\mathcal{C}_{1},\dots,\mathcal{C}_{k}\subset\mathbb{R}^{n}$.

\section{Methodology}\label{sec:method}

\subsection{Motivating example}\label{sec:motivation}

Here we provide a toy example that illustrates the definition of joint and individual structures of JIVE \citep{lock2013joint}, COBE \citep{zhou2016group} and AJIVE \citep{feng2018angle}, and how this definition does not account for partial sharing. We also illustrate how it is possible to have non-orthogonality between different structures (individual, joint, partially-shared), thus violating identifiability conditions for SLIDE \citep{gaynanova2019structural}.

\begin{ex}\label{ex:d3}
\normalfont
Consider $D=3$ with the signal matrices:
\[
\begin{aligned}
&\M_{1} =\begin{bmatrix}
1 & 1 & 1 \\
-1 & 1 & 0 \\
1 & 0 & -1 \\
-1 & 0 & 0 \\
1 & 0 & 0 \\
\end{bmatrix},\quad \M_{2} =\begin{bmatrix}
1 & 1 & 0 \\
-1 & 1 & 0 \\
1 & 0 & 1 \\
-1 & 0 & 0 \\
1 & 0 & -1 \\
\end{bmatrix}, \quad 
\M_{3} = \begin{bmatrix}
1 & 1 & 0 \\
-1 & 0 & -1 \\
1 & -1 & 0 \\
-1 & 0 & 2 \\
1 & 0 & 1 \\
\end{bmatrix}.
\end{aligned}        
\]
JIVE, COBE and AJIVE define the joint structure as the intersection of the column spaces 
$
\mathcal{C}(\M_{1})\cap\mathcal{C}(\M_{2})\cap\mathcal{C}(\M_{3})=\text{span}\left\{[1\quad -1\quad 1 \quad -1\quad 1]^\top\right\},
$
with the individual structures being the remaining signal:
\[
\begin{aligned}
&\text{Individual view 1:}\quad \mathcal{C}(\P_{\cap_{d=1}^{3}\mathcal{C}(\M_{d})}^\perp\M_{1}) = \text{span}\left\{[1\quad 1\quad 0 \quad 0\quad 0]^\top,\ [1\quad 0\quad -1 \quad 0\quad 0]^\top\right\},\\ 
&\text{Individual view 2:}\quad \mathcal{C}(\P_{\cap_{d=1}^{3}\mathcal{C}(\M_{d})}^\perp\M_{2}) = \text{span}\left\{[1\quad 1\quad 0 \quad 0\quad 0]^\top,\ [0\quad 0\quad 1 \quad 0\quad -1]^\top\right\}, \\
&\text{Individual view 3:}\quad \mathcal{C}(\P_{\cap_{d=1}^{3}\mathcal{C}(\M_{d})}^\perp\M_{3}) = \text{span}\left\{[1\quad 0\quad -1 \quad 0\quad 0]^\top,\ [0\quad -1\quad 0 \quad 2\quad 1]^\top\right\}.
\end{aligned}    
\]

However, these individual structures are not truly unique to each view: while their overall intersection is zero, they have non-trivial pairwise intersections. Concretely, it follows that (i)~$\text{span}\{[1\quad 1\quad 0 \quad 0\quad 0]^\top\}$ is shared by the 1st and 2nd views (but not the 3rd view); (ii) $\text{span}\{[1\quad 0\quad -1 \quad 0\quad 0]^\top\}$ is shared by the 1st and 3rd views (but not the 2nd view). Likewise, in many applications, it is reasonable to expect that certain signals are shared across a few views rather than all views. We refer to such signals as partially-shared. 
\end{ex}

Example~\ref{ex:d3} illustrates how many existing methods \citep{lock2013joint,zhou2016group,yang2016non,feng2018angle} treat partially-shared structures as individual. To consider partially-shared structures, we can apply these methods separately to each of the $\binom{D}{2}$ view pairs: if the estimated joint rank from any pair is larger than that from all $D$ views, it indicates the existence of a partially-shared structure between the corresponding pair. However, such approach significantly increases computational burden associated with corresponding methods and the resulting estimates across pairs are not guaranteed to be consistent with each other due to estimation errors.

To address these challenges,  we propose to define partially-shared and individual structures in terms of column spaces of signal matrices. As existing models, we define the joint structure as an intersection of all column spaces. In contrast to these methods, we further decompose each  $\mathcal{C}(\P_{\cap_{d=1}^{3}\mathcal{C}(\M_{d})}^\perp\M_{d})$ to incorporate partially-shared structures. Our key idea is to consider remaining signal structures after accounting for joint structure. 
More precisely, the partially-shared signal structures (rigorously defined in Definition \ref{def:spaces}) in Example~\ref{ex:d3} are specified as 
\[
\begin{aligned}
&J(1,2) = \mathcal{C}(\P_{\cap_{d=1}^{3}\mathcal{C}(\M_{d})}^\perp\M_{1})\cap\mathcal{C}(\P_{\cap_{d=1}^{3}\mathcal{C}(\M_{d})}^\perp\M_{2})=\text{span}\left\{[1\quad 1\quad 0 \quad 0\quad 0]^\top\right\}, \\
&J(1,3) =\mathcal{C}(\P_{\cap_{d=1}^{3}\mathcal{C}(\M_{d})}^\perp\M_{1})\cap\mathcal{C}(\P_{\cap_{d=1}^{3}\mathcal{C}(\M_{d})}^\perp\M_{3})=\text{span}\left\{[1\quad 0\quad -1 \quad 0\quad 0]^\top\right\}, \\
&J(2,3) = \{\mathbf{0}\}.
\end{aligned}
\]
By considering the remaining signals, we redefine the individual structures in Example \ref{ex:d3} as 
\[
\begin{aligned}
&\text{(proposed) individual view 1:}\quad I(1) = \{\mathbf{0}\},\\ 
&\text{(proposed) individual view 2:}\quad I(2) = \text{span}\left\{\ [0\quad 0\quad 1 \quad 0\quad -1]^\top\right\},\\
&\text{(proposed) individual view 3:}\quad I(3) = \text{span}\left\{\ [0\quad -1\quad 0 \quad 2\quad 1]^\top\right\}.
\end{aligned}    
\]
In comparison to COBE, JIVE and AJIVE, our approach models the individual structures that are indeed view-specific after taking into account both partially-shared structures in Example~\ref{ex:d3}. Our signal structure modeling is discussed with more details in Section \ref{subsec:def}.

At the same time, SLIDE model \citep{gaynanova2019structural} also allows partial-sharing but requires orthogonality between different types of structures to guarantee uniqueness. However, in Example~\ref{ex:d3}, the individual and partially-shared structures are not orthogonal to each other, that is $J(1,2)\not\perp J(1,3)$, $J(1,2)\not\perp I(3)$, $J(1,3)\not\perp I(2)$ and $I(2)\not\perp I(3)$. This implies that SLIDE's identifiability conditions are violated, and the resulting parametrization of SLIDE model is not unique, leading to potential difficulties in estimation and interpretation. This is illustrated in Example \ref{ex:slide} in Appendix \ref{supp:method}.

\subsection{Proposed definition of hierarchical signal structure}\label{subsec:def}

Motivated by Example \ref{ex:d3}, we propose to define joint, partially-joint, and individual structures via hierarchical levels. When $D=3$, we consider three hierarchical levels: 
\[
\text{Level 1}:\ \left\{(1,2,3)\right\} ,\ \text{Level 2}:\ \left\{(1,2),(1,3),(2,3)\right\},\ \text{Level 3}:\ \left\{(1),(2),(3)\right\},
\]
where the integers indicate the views. Levels 1, 2 and 3 correspond to the joint structure, partially-shared structures between $\binom{3}{2}$ distinct view pairs, and the three individual structures, respectively. The structure at each level is defined recursively based on the previous level, starting with the intersection of all column-spaces $\mathcal{C}(\M_d)$ at Level 1. Figure~\ref{def:spaces} provides an illustration of the proposed definition as applied to Example~\ref{ex:d3}, which we formalize below.

\begin{figure}[H]\label{fig1}
\centering
\includegraphics[scale = 0.55]{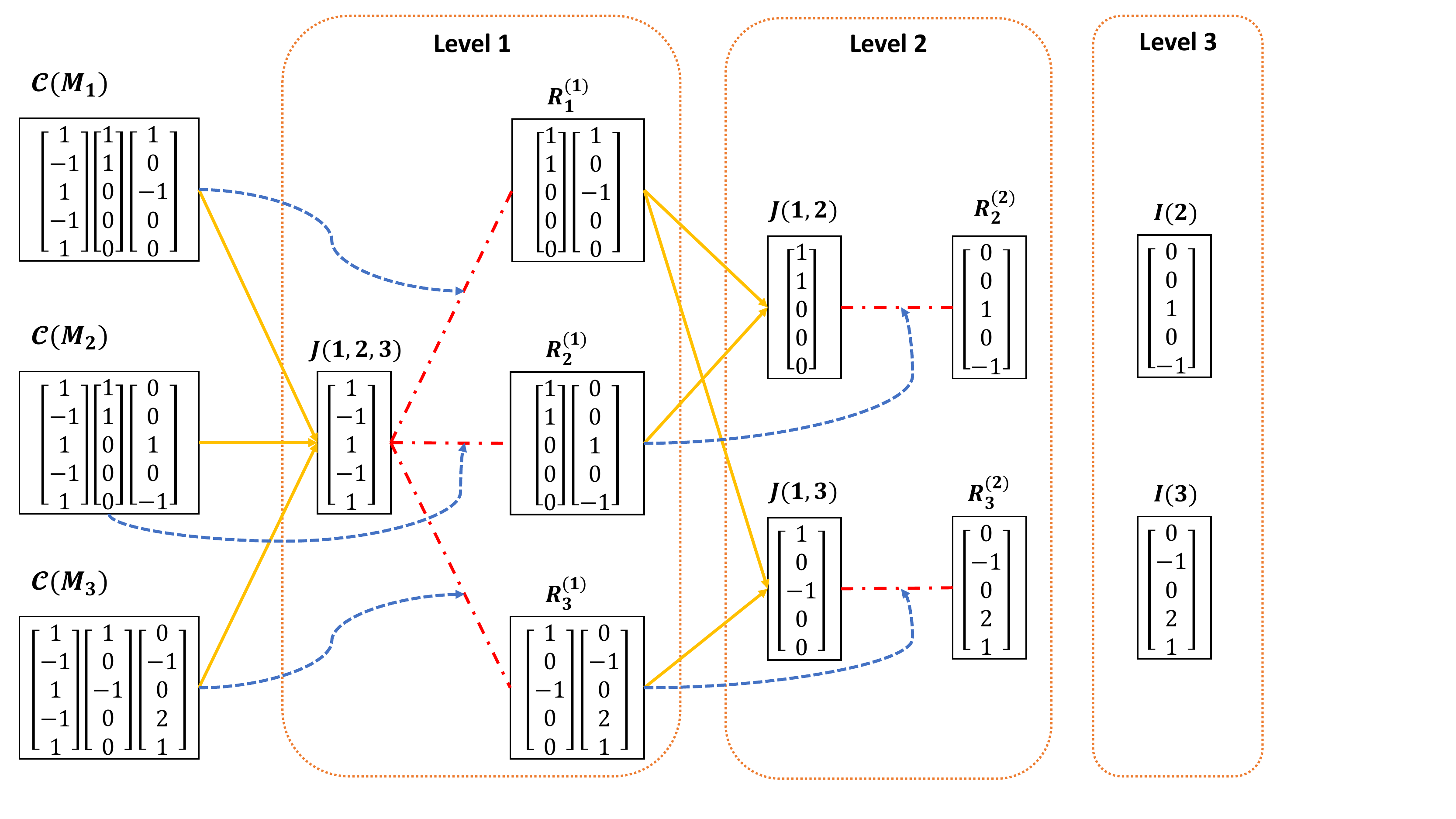}
\caption{Illustration of Definition~\ref{def:spaces} applied to the signals in Example~\ref{ex:d3}. The solid arrows (in yellow) indicate the corresponding intersection of column-spaces (joint structures). The red dash-dotted lines (in red) together with the dotted arrows (in blue) indicate the remainder terms after accounting for joint structures. Observe the missing $J(2, 3)$ (due to zero intersection or $R_2^{(1)}$ and $R_3^{(1)}$), and missing $R_1^{(2)}$, $I(1)$ (due to $R_1^{(1)}$ being fully captured by $J(1, 2)$ and $J(1, 3)$)}\label{fig:flow}
\end{figure}

At Level 1, the joint structure is defined as $J(1,2,3) = \bigcap\limits_{d=1}^{3} \mathcal{C}(\M_{d})$. We use $R_{d}^{(1)}= S_{J(1,2,3)}^\perp(\M_{d})$ to represent the remaining signal in each view after accounting for the joint structure.
At Level 2, we use the signal remaining from Level 1 to define the joint structures across each view pair. Formally, $J(l,m) = R_{l}^{(1)} \cap R_{m}^{(1)}$ for $(l,m) = (1,2),(1,3),(2,3)$. For example, $J(1,2)$ is shared across views 1 and 2, is not present in view 3, and  by construction is orthogonal to the globally-shared $J(1, 2, 3)$. We use $R_{d}^{(2)}$ to define the remaining signal in each view after accounting for structures in Levels 1-2. That is, $R_{1}^{(2)}=S_{Q_{1}^{(2)}}^\perp(R_{1}^{(1)})$, $ R_{2}^{(2)}=S_{Q_{2}^{(2)}}^\perp(R_{2}^{(1)})$ and $R_{3}^{(2)}=S_{Q_{3}^{(2)}}^\perp(R_{3}^{(1)})$
where  
$
Q_{1}^{(2)}=\text{span}\{J(1,2),J(1,3)\}$, $Q_{2}^{(2)}=\text{span}\{J(1,2),J(2,3)\}$ and $Q_{3}^{(2)}=\text{span}\{J(1,3),J(2,3)\}.
$
At Level 3, only singletons remain, with individual structure for view $d$ being the remainder $I(d) = R_{d}^{(2)}$ since it is already adjusted for both joint and partially-shared structures at the previous two levels. 

For general $D$, there are in total $D$ hierarchical levels. For $k\in\{1,\dots, D\}$, the $(D-k+1)$-th level corresponds to the structures shared across $k$ views, and there are $\binom{D}{k}$ distinct view combinations $(i_{1},\dots,i_{k})$.
The key idea is to recursively decompose the remaining signal from the previous level into the shared structures specified at the current level and the remaining signal.
Below, we provide the full definition of joint, partially-joint and individual structures.

\begin{my_def}\label{def:spaces}
\normalfont
At Level 1 $ = \{(1, \dots, D)\}$ the corresponding joint structure and the remaining signals are defined as
$
J(1,\dots,D) = \bigcap\limits_{d=1}^{D} \mathcal{C}(\M_{d})$ and  $R_{d}^{(1)} = 
S_{J(1,\dots,D)}^\perp(\M_{d}).
$

At Level $(D-k+1)$ for $2\leq k\leq (D-1)$, the partially-shared signal for each distinct view combination $(i_{1},\dots,i_{k})$ is the intersection of the remaining signals from the previous level:
\[
J(i_1,\dots,i_k) = \bigcap\limits_{d=i_{1}}^{i_{k}} R_{d}^{(D-k)},
\]
We call $J(i_1,\dots,i_k)$ the $k$-way partially-shared structure common to views $i_{1},\dots,i_{k}$. Let $Q_{d}^{(D-k+1)}$ be the space spanned by all $\binom{D-1}{k-1}$ distinct $J(i_1,\dots,i_k)$ with $d\in\{i_1,\dots,i_k\}$ (the $k$-way partially-shared spaces associated with view $d$). The remaining signal in view $d$ is
\[
R_{d}^{(D-k+1)}=S_{Q_{d}^{(D-k+1)}}^\perp(R_{d}^{(D-k)}).
\]
At Level $D$, the individual structure is defined as $I(d)=R_{d}^{(D-1)}$.
\end{my_def}

The main advantage of Definition \ref{def:spaces} is that each type of structure is defined in terms of column space, instead of relying on matrix representations (via decomposition and factorization) that existing methods \citep{lock2013joint, zhou2016group, yang2016non, feng2018angle, gaynanova2019structural} use.
As illustrated below, this facilitates a clean and non-restrictive definition of partially-shared structures. Take $D=3$ views as an example. Definition \ref{def:spaces} decomposes $\mathcal{C}(\M_{d})$ as

\begin{equation}\label{decomp:spaces}
\begin{aligned}
&\mathcal{C}(\M_{1}) = \underbrace{J(1,2,3)}_{\text{Level 3}}+\underbrace{\text{span}\{J(1,2),J(1,3)\}}_{\text{Level 2}}+\underbrace{I(1)}_{\text{Level 1}} \\
&\mathcal{C}(\M_{2}) = \underbrace{J(1,2,3)}_{\text{Level 3}}+\underbrace{\text{span}\{J(1,2),J(2,3)\}}_{\text{Level 2}}+\underbrace{I(2)}_{\text{Level 1}}, \\
&\mathcal{C}(\M_{3}) = \underbrace{J(1,2,3)}_{\text{Level 3}}+\underbrace{\text{span}\{J(1,3),J(2,3)\}}_{\text{Level 2}}+\underbrace{I(3)}_{\text{Level 1}}. \\
\end{aligned}        
\end{equation}
In the column-space representation, JIVE, COBE and AJIVE
correspond to the decomposition $\mathcal{C}(\M_{d}) =  J(1,2,3) + I_{\text{JIVE}}(d)$, where $I_{\text{JIVE}}(d)=\mathcal{C}(\P_{J(1,2,3)}^\perp\M_{d})$ denotes their individual structure for $d=1,2,3$. In contrast, our formulation makes a further decomposition of $I_{\text{JIVE}}(d)$, making partially-shared structures explicit.

Furthermore, by construction, Definition \ref{def:spaces} allows the spaces within the same level be non-orthogonal to each other: (i) $J(1,2)$, $J(1,3)$ and $J(2,3)$ at Level 2; (ii) $I(1)$, $I(2)$ and $I(3)$ at Level 3. Additionally, the structures that do not contain the same view can be non-orthogonal such as (i) $J(1,2)$ and $I(3)$; (ii) $J(1,3)$ and $I(2)$; (iii) $J(2,3)$ and $I(1)$. Since our approach does not impose extra orthogonality conditions and is based directly on the column spaces of signal matrices, it overcomes identifiability issues of SLIDE model. 

\subsection{Hierarchical rank constraints}\label{subsec:hnnr}

Based on Definition \ref{def:spaces}, one possibility is to use matrix factorization for signal estimation, that is to decompose each $\M_{d}$ into a sum of several matrices that represent each signal structure. The scores and loadings factorization of each structure can be used for estimation, which is pursued by many existing methods for data integration \citep{lock2013joint, feng2018angle, gaynanova2019structural}. In our framework, however, it is unclear how to adopt the scores and loadings factorization while preserving the potential non-orthogonality of the signal structures. Another possibility is to consider signal estimation as optimization problem with explicit subspace constraints \citep{yuan2022double}. However, in presence of multiple types of structures, this leads to a highly non-trivial non-convex, manifold optimization. Here, we propose a new strategy for direct signal estimation that avoids both matrix factorization, and subspace constraints, leading to a convex optimization problem. Our key idea is to introduce hierarchical nuclear norm penalty, which is motivated by the rank constraints imposed on the matrices at each hierarchical levels discussed in this section.  

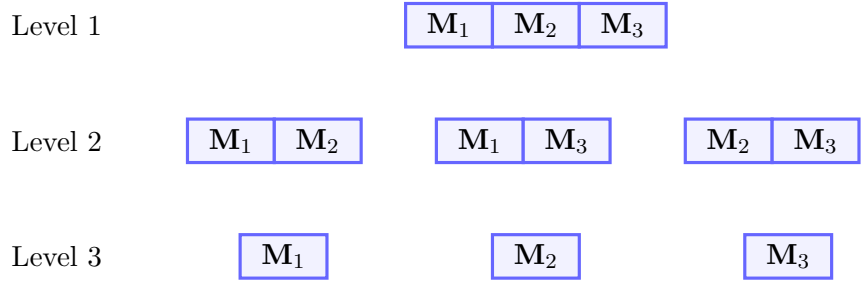
\begin{figure}
    \begin{center}
    \begin{tikzpicture}[
    square/.style = {draw=blue!60, fill=blue!5, very thick, 
                 minimum height=1.2em, minimum width=3em, outer sep=0pt}]
                 
    \node[rectangle, draw = white] (L1) {Level 1};
    \node[square, right=3.9cm of L1]   (M1_3) {$\M_{1}$};
    \node[square, right=0cm of M1_3] (M2_3) {$\M_{2}$};
    \node[square, right=0cm of M2_3] (M3_3) {$\M_{3}$};
    
    \node[rectangle, draw = white, below = of L1] (L2) {Level 2};
    \node[square, right=of L2]   (M1_2) {$\M_{1}$};
    \node[square, right=0cm of M1_2] (M2_2) {$\M_{2}$};
    
    \node[square, right=of M2_2]   (M11_2) {$\M_{1}$};
    \node[square, right=0cm of M11_2] (M3_2) {$\M_{3}$};
    
    \node[square, right=of M3_2]   (M22_2) {$\M_{2}$};
    \node[square, right=0cm of M22_2] (M33_2) {$\M_{3}$};    
    
    \node[rectangle, draw = white, below = of L2] (L3) {Level 3};
    \node[square, right=1.7cm of L3]   (M1_1) {$\M_{1}$};    
    \node[square, right=2.2cm of M1_1]   (M2_1) {$\M_{2}$};    
    \node[square, right=2.2cm of M2_1]   (M3_1) {$\M_{3}$};    
    \end{tikzpicture}
    
    \end{center}
    \caption{A diagram showing the matrices in $\mathcal{H}_{D-k+1}(\M)$ with $\M = [\M_{1}\ \M_{2}\ \M_{3}]$}
    \label{fig:hnn_matrices}
\end{figure}

Specifically, we propose to take advantage of the constraint on the ranks of the matrices at each hierarchical levels. Let $\mathcal{I}_{D-k+1}$ be the $(D-k+1)$-th hierarchical level as in Section \ref{subsec:def}. Given the total signal $\M = [\M_{1}\ \dots \M_{D}]$, define the set $\mathcal{H}_{D-k+1}(\M)$ as
$\mathcal{H}_{D-k+1}(\M)=\left\{ \M_{(i_1,\dots,i_k)}: (i_1,\dots,i_k)\in \mathcal{I}_{D-k+1} \right\}$,
where $\M_{(i_1,\dots,i_k)}$ denotes the submatrix of $\M$ formed by concatenating each $\M_{d}$ for $d\in \left\{i_1,\dots,i_k\right\}$. That is, $\mathcal{H}_{D-k+1}(\M)$ is the collection of $\binom{D}{k}$ distinct submatrices of $\M$ corresponding to each vector of $\mathcal{I}_{D-k+1}$. Figure \ref{fig:hnn_matrices} shows the matrices of each $\mathcal{H}_{D-k+1}(\M)$ when $D=3$. As shown in Section~\ref{sec:motivation}, the existing models' definition of joint structure \citet{lock2013joint, zhou2016group, feng2018angle} relies on non-trivial intersection of column spaces of signal matrices, which implies rank constraint:
\begin{equation*}
r_{123} = \rank([\M_1\ \M_2\ \M_3])  < \rank(\M_1) + \rank(\M_2) + \rank(\M_3) = r_1 + r_2 + r_3.
\end{equation*}
When estimating $\bM_d$ from noisy $\X_d$ in model~\eqref{eq:ourmod}, these methods effectively impose the low-rank constraints with some $r_d\ll\min(n,p_d)$, $d=1, 2, 3$, $r_{123} \ll \min(n, \sum_{d=1}^3p_d)$ so that
\begin{equation}\label{eq:rankD3}
 \rank([\M_1\ \M_2\ \M_3])\leq r_{123},\ \rank(\bM_d)\leq r_d, \ d=1, 2, 3.
\end{equation}
Choosing ranks such that $r_{123}< \sum_{d=1}^3r_d$ leads to non-trivial intersections in column spaces.
Using the intuition from \eqref{eq:rankD3}, we introduce the hierarchical rank constraints which are imposed on the ranks of all matrices in $\mathcal{H}_{D-k+1}(\M)$ for all hierarchical levels. Using $D=3$ example in Figure \ref{fig:hnn_matrices}, the constraints become
\begin{equation}\label{eq:hranks}
\begin{aligned}
&\rank([\M_{1}\ \M_{2}\ \M_{3}])\leq r_{123},\ \rank([\M_{1}\ \M_{2}])\leq r_{12},\ \rank([\M_{1}\ \M_{3}])\leq r_{13},\\  
&\rank([\M_{2}\ \M_{3}])\leq r_{23},\ \rank(\M_{1})\leq r_{1},\ \rank(\M_{2})\leq r_{2},\ \rank(\M_{3})\leq r_{3}.
\end{aligned}
\end{equation}
The constraints for $D>3$ are similar. Various combinations of ranks in \eqref{eq:hranks} lead to intersection of column spaces with varying dimensions, allowing a wide range of signal structures. However, direct use of constraints \eqref{eq:hranks} in estimations leads to a non-convex, NP-hard optimization problem \citep{candes2009exact,mazumder2010spectral}. Instead, in Section \ref{subsec:HNN} we propose to replace each constraint in \eqref{eq:hranks} with corresponding nuclear norm penalty.

\subsection{Hierarchical nuclear norm penalization}\label{subsec:HNN}

Nuclear norm penalization \citep{bach2008consistency, candes2009exact, negahban2011estimation} is commonly used for the estimation of the low-rank signal $\M_{d}$ in single-view settings since nuclear norm is a convex relaxation to rank leading to convex optimization problem
\begin{equation}\label{single}
\minimize_{\M_{d}} \left\{ \frac{1}{2}\|\X_{d}-\M_{d}\|_{F}^{2} + \lambda_{d}\|\M_{d}\|_{*} \right\},    
\end{equation}
where $\lambda_{d}\geq0$ is a tuning parameter. Parallel to $\ell_{1}$-norm penalization for sparse vector estimation, the nuclear norm penalty in \eqref{single} encourages sparse singular values for low-rank signal estimation. The solution to \eqref{single} can be written in closed-form via the proximal operator of the nuclear norm \citep{polson2015proximal}:
\begin{equation}\label{sft}
S(\X_{d},\lambda_{d}) = \sum_{i=1}^{\text{min}(n,p_{d})}(\sigma_{i}(\X_{d}) - \lambda_{d})_{+}u_{i}v_{i}^\top\in\mathbb{R}^{n\times p_{d}}, 
\end{equation}
where $x_{+} = \text{max}(x,0)$ and $u_{i}\in\mathbb{R}^{n\times1}$ and $v_{i}^\top\in\mathbb{R}^{1\times p_{d}}$ are the left and right singular vectors corresponding to $\sigma_{i}(\X_{d})$, respectively. That is, $S(\X_{d},\lambda_{d})$ is the matrix resulting from soft-thresholding the singular values of $\X_{d}$ with cutoff $\lambda_{d}$.  The larger is the value of $\lambda_{d}$, the stronger is the penalty, and subsequently the lower is the rank of the resulting $S(\X_{d},\lambda_{d})$.

\begin{algorithm}[!t]
\caption{Dual block-coordinate forward-backward algorithm for \eqref{opt:3views}}\label{algo:dbfb}
\begin{algorithmic}

\STATE \text{Input}: $\X = [\X_{1}\ \X_{2}\ \X_{3}]$, $\lambda,\lambda_{12},\lambda_{13},\lambda_{23},\lambda_{1},\lambda_{2},\lambda_{3}\geq0$, $\epsilon>0$, $\gamma\in(0,2)$, $t=1$.

\STATE Initialize with $\M^{(0)} = \X$, $\D_{d}^{(0)}=\mathbf{0}$ for $d=1,2,3$ and 
\STATE \hspace{2.5cm} $\D_{kl}^{(0)}=\mathbf{0}$ for $(k,l) = (1,2),(1,3),(2,3)$.
    
\REPEAT
    \STATE $\D_{d}^{(t)} = \D_{d}^{(t-1)} + \gamma\M_{d}^{(t-1)} - S\left( \D_{d}^{(t-1)} + \gamma\M_{d}^{(t-1)}, \lambda_{d} \right)$,  
    \STATE $\D_{kl}^{(t)} = \D_{kl}^{(t-1)} + \gamma\M_{kl}^{(t-1)} - S\left( \D_{kl}^{(t-1)} + \gamma\M_{kl}^{(t-1)}, \lambda_{kl} \right)$, 
    \STATE $\D^{(t)} = \D^{(t-1)} + \gamma\M^{(t-1)} - S\left( \D^{(t-1)} + \gamma\M^{(t-1)}, \lambda \right)$,
    \STATE $\begin{aligned}
    \M^{(t)}= &\M^{(t-1)} - \left[(\D_1^{(t)}-\D_1^{(t-1)}) \quad (\D_2^{(t)}-\D_2^{(t-1)})\quad (\D_3^{(t)}-\D_3^{(t-1)})\right]\\
    -&\left[(\D_{12}^{(t)}-\D_{12}^{(t-1)})\ \mathbf{0}_{n\times p_{3}}\right] 
    -\left[(\D_{13,1}^{(t)}-\D_{13,1}^{(t-1)})\ \mathbf{0}_{n\times p_{2}}\ (\D_{13,2}^{(t)}-\D_{13,2}^{(t-1)})\right] \\ 
    -&\left[\mathbf{0}_{n\times p_{1}}\ (\D_{23}^{(t)}-\D_{23}^{(t-1)})\right] - (\D^{(t)}-\D^{(t-1)})
    \end{aligned}$
    \STATE
    \STATE $t = t+1$
\UNTIL{$\|\M^{(t)}-\M^{(t-1)}\|_{F}<\epsilon$.}

\STATE \text{Return}: $\widehat{\M}= [\widehat{\M}_{1}\ \widehat{\M}_{2}\ \widehat{\M}_{3}]=\M^{(t)}$
\end{algorithmic}
\end{algorithm}

Motivated by the above hierarchical rank constraints~\eqref{eq:hranks} in the multi-view case, we propose the hierarchical nuclear norm (HNN) penalty:
\begin{equation}\label{eq:HNNpenalty}
\bm{\lambda}
\text{HNN}(\M,\bl) = \sum\limits_{\M_{(i_1,\dots,i_k)}\in \bigcup\limits_{k=1}^{D}\mathcal{H}_{D-k+1}(\M)} \lambda_{i_{1}\dots i_{k}} \|\M_{(i_1,\dots,i_k)}\|_{*},
\end{equation}
where $\M_{(i_1,\dots,i_k)}$ is the submatrix of $\M$ formed by concatenating $\M_{d}$ for $d\in \left\{i_1,\dots,i_k\right\}$, and $\lambda_{i_{1}\dots i_{k}}\geq0$ is the corresponding tuning parameter that controls the amount of shrinkage applied to the singular value of each matrix in $\bigcup\limits_{k=1}^{D}\mathcal{H}^{(k)}(\M)$. By encouraging sparse singular values of matrices at each hierarchical level, \eqref{hnn:opt} naturally induces different signal structures due to the correspondence between rank and the number of non-zero singular values.

The corresponding joint hierarchical nuclear norm penalization problem is
\begin{equation}\label{hnn:opt}
\argmin_{\M_{1},\dots,\M_{D}} \left\{ \frac{1}{2}\|\X_{d}-\M_{d}\|_{F}^{2} + \text{HNN}(\M,\bl) \right\},  
\end{equation}
with the solution path covering a wide range of models with different configurations of hierarchical ranks. The choice of the tuning parameters will be discussed in Section \ref{subsec:bcv}.
For an example of $D=3$, let $\M_{kl}=[\M_{k}\ \M_{l}]\in\mathbb{R}^{n \times (p_{k}+p_{l})}$, then \eqref{hnn:opt} reduces to 
\begin{equation}\label{opt:3views}
\minimize_{\M_{1},\M_{2},\M_{3}}  \biggl\{ \frac{1}{2}\sum_{d=1}^{3}\left\|\X_{d}-\M_{d}\right\|_{F}^{2} + \sum_{d=1}^{3}\lambda_{d}\|\M_{d}\|_{*}+\sum_{k,l=1,k<l}^{3}\lambda_{kl}\|\M_{kl}\|_{*}  +\lambda\|\M\|_{*} \biggr\}.   
\end{equation}

Note that our approach is not based on matrix factorizations,
which would lead to a non-convex optimization. Instead, \eqref{hnn:opt} is a convex optimization problem, which can be efficiently solved by the dual block-coordinate forward-backward algorithm \citep{abboud2017dual}.
We specialize this general algorithm to solving \eqref{hnn:opt}.
The pseudo-code is presented in Algorithm \ref{algo:dbfb}.
The detailed derivation is deferred to Appendix \ref{supp:method}.
One important finding is that the proximal operator of the hierarchical nuclear norm can be evaluated efficiently because the norm is a sum of convex functions with linear operators.

\subsection{Refitting procedure}\label{subsec:refitting}

Like many penalized estimation methods, it is well-known that the nuclear norm penalization causes shrinkage bias \citep{chen2013reduced, josse2016adaptive}. In particular, since it penalizes both small and large non-zero singular values by equal amount, the so-called overshrinkage phenomenon could occur where non-zero singular values are shrunken too much. This causes inaccurate estimation of signal as the non-zero singular values are likely to be estimated smaller than the original values. 

Since our hierarchical nuclear norm penalty consists of the nuclear norms of multiple matrices, our approach is also affected by such overshrinkage phenomenon. Thus, we adopt a simple refitting step to ``un-shrink'' our estimate. The key idea is to extract estimated low-rank column space from preliminary $\widehat{\M}_{d}$ from Algorithm \ref{algo:dbfb}, and then construct $\widehat{\M}_d^{\text{refit}}$ with the given column space without penalization of singular values. This idea is formalized in Algorithm \ref{algo:refit}. Since the refitting step corresponds to solving the least-squares problem \eqref{opt:refit} with a closed-form solution, it does not require a significant amount of computation. As illustrated in Figure \ref{fig:refit} of Appendix \ref{supp:method}, while the singular values of each $\widehat{\M}_{d}$ are smaller than the true values, our refitting procedure is effective in correcting the bias.

\begin{algorithm}[!t]
\caption{Refitting procedure for each view when $D=3$}\label{algo:refit}
\begin{algorithmic}
\STATE \text{Input}: $\X = [\X_{1}\ \X_{2}\ \X_{3}]$ and $\widehat{\M}= [\widehat{\M}_{1}\ \widehat{\M}_{2}\ \widehat{\M}_{3}]$ from Algorithm \ref{algo:dbfb}
\FOR{$d=1,2,3$}
\STATE (i) Obtain the left singular matrix $\widehat{\U}_{d}\in\mathbb{R}^{n\times r_{d}}$ of $\widehat{\M}_{d}$ corresponding to its $r_{d}$ non-zero 
\STATE \hspace{.65cm}singular values
\STATE (ii) Calculate
\begin{equation}\label{opt:refit}
\widehat{\V}_{d}^\top=\widehat{\U}_{d}^\top\X_{d}=\argmin_{\V_{d}^\top} \|\X_{d}-\widehat{\U}_{d}\V_{d}^\top\|_{F}^{2}\in\mathbb{R}^{r_{d}\times p_{d}}.
\end{equation}
\ENDFOR
\STATE \text{Return}: $\widehat{\M}^{\text{refit}}= [\widehat{\U}_{1}\widehat{\V}_{1}^\top\ \widehat{\U}_{2}\widehat{\V}_{2}^\top\ \widehat{\U}_{3}\widehat{\V}_{3}^\top]$
\end{algorithmic}
\end{algorithm}

\subsection{Selection of tuning parameters}\label{subsec:bcv}

The tuning parameters in the HNN penalty in~\eqref{eq:HNNpenalty} need to be properly chosen for the good performance of our approach. If the nuclear norms are over-penalized, there is a degradation in signal estimation performance due to underestimated ranks (which the above refitting procedure cannot correct). On the other hand, if the tuning parameters are too small, the ranks are overestimated, and the estimated signals contain noise. One clear difficulty in selecting tuning parameters for~\eqref{hnn:opt} is the large number of total parameters. For instance, even a modest $D=3$ case \eqref{opt:3views} requires 7 parameters. Tuning all these parameters freely via cross-validation or a model selection criterion is computationally prohibitive. 

We propose to reduce the total number of tuning parameters from $2^D-1$ to only $D$ by taking advantage of the Stein's Unbiased Risk Estimates (SURE) \citep{candes2013unbiased,josse2016adaptive}.
The SURE formulation for single-view nuclear-norm minimization allows to derive a closed form of optimal $\lambda_{d}^{\text{SURE}}$ to use in $S(\X_{d},\lambda_{d})$~\eqref{sft} as a minimizer of SURE criterion, see \cite{candes2009exact} for the closed-form of $\lambda_{d}^{\text{SURE}}$ and derivation details. While SURE formulation for HNN minimization is intractable, we propose to take advantage of closed-form $\lambda_{d}^{\text{SURE}}$ in single matrix case to determine the relative weights that should be assigned to each matrix within each hierarchical level $\mathcal{H}_{D-k+1}(\M)$. Specifically, let $\lambda_{d}^{\text{SURE}}$ and $\lambda_{kl}^{\text{SURE}}$ be the SURE tuning parameters for $S(\X_{d},\lambda_{d})$ and $S(\X_{kl},\lambda_{kl})$, respectively, where $\X_{kl}=[\X_{k}\ \X_{l}]$. We propose to reparametrize penalty in \eqref{opt:3views} as
\begin{equation}\label{pen:ratio}
\tau\sum_{d=1}^{3} \omega_{d}^{\text{SURE}}\|\M_{d}\|_{*} + \kappa\sum_{k,l=1,k<l}^{3}\omega_{kl}^{\text{SURE}}\|\M_{kl}\|_{*}, +\lambda\|\M\|_{*} 
\end{equation}    
where $\tau,\kappa>0$ control the overall shrinkage amount for the individual and pairwise matrices, respectively, and the relative weights are chosen as
\begin{equation}\label{ratio:tune}
\omega_{d}^{\text{SURE}} = \frac{\lambda_{d}^{\text{SURE}}}{\lambda_{1}^{\text{SURE}} + \lambda_{2}^{\text{SURE}} + \lambda_{3}^{\text{SURE}}}, \quad \omega_{kl}^{\text{SURE}} = \frac{\lambda_{kl}^{\text{SURE}}}{\lambda_{12}^{\text{SURE}} + \lambda_{13}^{\text{SURE}} + \lambda_{23}^{\text{SURE}}}.    
\end{equation}
That is, we form a convex combination of the nuclear norms within each hierarchical level in the penalty \eqref{pen:ratio}. The approach for $D>3$ is similar. By design, $\tau$, $\kappa$ and $\lambda$ control the amount of penalization across different hierarchical levels, while the weights \eqref{ratio:tune} induce (potentially) differential shrinkage within the same level based on SURE. Since the weights are closed form, only three parameters $(\tau,\kappa,\lambda)$ remain to be tuned.

To choose the remaining $D$ tuning parameters, we adapt the bi-cross-validation (BCV) approach \citep{owen2009bi}, similar approach is used in SLIDE \citep{gaynanova2019structural}. The original BCV is designed for rank selection in the single-view case, and can be viewed as matrix extension of $k$-fold cross-validation. The $(j\times k)$-fold BCV splits the rows and the columns of $\X$ into $j$ and $k$ blocks, respectively. The submatrix corresponding to each block is hold out as test data, with the remaining submatrices used to fit the model of specified rank and make predictions on the test. 

To illustrate our application of the BCV, we focus on the $2\times2$ folds of $\X$ such that  
\begin{equation}\label{bcv:split}
\X=[\X_{1}\ \X_{2}\ \X_{3}]=
\delimitershortfall=0pt
\setlength{\dashlinegap}{2pt}
\begin{bmatrix}[cc:cc:cc]
\X_{1}^{1,1} & \X_{1}^{1,2} &  \X_{2}^{1,1} & \X_{2}^{1,2} &  \X_{3}^{1,1} & \X_{3}^{1,2} \\
\X_{1}^{2,1} & \X_{1}^{2,2} &  \X_{2}^{2,1} & \X_{2}^{2,2} &  \X_{3}^{2,1} & \X_{3}^{2,2} \\
\end{bmatrix}    
\end{equation}
where $\X_{d}^{j,k}$ denotes the sub-matrix of $\X_{d}$ corresponding to the $j$-th row fold and $k$-th column fold. Algorithm \ref{algo:bcv} summarizes the corresponding BCV procedure. The main difference with original BCV is that the splits are distributed equally across the views. Suppose we hold out $\X_{d}^{j,k}$ for all $d=1,2,3$. Given the tuning parameters $(\tau_{l},\kappa_{l},\lambda_{l})$, we obtain the estimates $\widehat{\M}_{-j,-k,l}^{\text{refit}}$ based on the submatrix of $\X$ that shares no rows or columns with held-out $\X_{j,k}^{(d)}$ (this submatrix is denoted by $\X_{-j,-k}^{(d)}$ in Algorithm \ref{algo:bcv}). We then use both $\X_{d}^{j,-k}$ (sharing only the rows with $\X_{d}^{j,k}$) and $\X_{d}^{-j,k}$ (sharing only the columns with $\X_{d}^{j,k}$) to evaluate the prediction error $\text{BCVErr}_{j,k,l}$ \eqref{bcverr} for $\X_{d}^{j,k}$ as in \cite{owen2009bi}. Given the grid of tuning parameters $\left\{(\tau_{l}, \kappa_{l}, \lambda_{l}) \right\}_{l=1}^{L}$, we calculate the average BCV error $\text{Avg\_Err}_{l} = 4^{-1}\sum_{j,k=1}^{2}\text{BCVErr}_{j,k,l}$ for combination $l$, and choose optimal $(\tau^{*}, \kappa^{*}, \lambda^{*})$ based on 1 standard error rule \citep{hastie2009elements}, see Appendix \ref{supp:method} for more details.

\begin{algorithm}[!t]
\begin{algorithmic}
\caption{Proposed $2\times 2$ BCV procedure for evaluation of prediction errors}\label{algo:bcv}
\STATE Input: $\X$ with random split as \eqref{bcv:split} and $\left\{(\tau_{l}, \kappa_{l}, \lambda_{l}) \right\}_{l=1}^{L}$
\FOR{$j,k=1,2$}
\STATE Calculate $\Z_{d}^{-j,-k}$ by column-centering and scaling $\X_{d}^{-j,-k}$ so that $\|\Z_{d}^{-j,-k}\|_{F}=1$  
\FOR{$l=1,2,\dots,L$}
\STATE (i) Obtain $\widehat{\M}_{-j,-k,l}^{\text{refit}}=[\widehat{\M}_{1, -j,-k,l}^{\text{refit}}\ \widehat{\M}_{2, -j,-k,l}^{\text{refit}}\ \widehat{\M}_{3, -j,-k,l}^{\text{refit}}]$ by back-scaling and \STATE back-centering the resulting matrix from Algorithm \ref{algo:refit} with $\Z_{d}^{-j,-k}$ and $(\tau_{l}, \kappa_{l}, \lambda_{l})$.
\STATE (ii) Evaluate the average BCV error such that 
\begin{equation}\label{bcverr}
\text{BCVErr}_{j,k,l}=\frac{1}{3}\sum_{d=1}^{3}\frac{1}{\Big\|\X_{d}^{j,k}\Big\|_{F}^{2}}{\left\|\X_{d}^{j,k}-\X_{d}^{j,-k}  \{\widehat{\M}_{d, -j,-k,l}^{\text{refit}}\}^{+} \X_{d}^{-j,k}\right\|_{F}^{2}},
\end{equation}
\STATE where the superscript $+$ denotes the pseudo-inverse of the corresponding matrix.
\ENDFOR
\ENDFOR
\STATE Return: $\{\text{BCVErr}_{j,k,l}:j,k=1,2\}_{l=1}^{L}$ 
\end{algorithmic}
\end{algorithm}

\section{Simulation studies}\label{sec:sim}

We conduct simulations to evaluate the performance of our approach. We label our approach as HNN and select its tuning parameters as described in Section~\ref{subsec:bcv}. To investigate the effect of our tuning parameter selection, we also consider our estimator with the smallest error \eqref{sfb} and label it as HNN\_best. For comparison, we further implement the other competing methods as follows: (a) JIVE \citep{lock2013joint} with the default permutation method for rank selection and true ranks (labeled as JIVE\_true); (b) AJIVE \citep{feng2018angle}; (c) SLIDE \citep{gaynanova2019structural}; (d) BIDIFAC \citep{park2020integrative} (e) BIDIFAC+ \citep{lock2022bidimensional}. Additional implementation details are discussed in Appendix \ref{supp:sim}. The R code for all analyses can be found at \url{http://github.com/sangyoonstat/HNN\_paper}.

We simulate each data matrix $\X_{d}$ via the model $\X_{d} = \M_{d} + \E_{d},$ where each element of the noise matrix $\E_{d}$ is generated independently from $N(0,\sigma_{d}^{2})$. The noise level $\sigma_{d}$ is set to satisfy the chosen signal-to-noise ratio (SNR), where $\mbox{SNR}=\|\M_{d}\|_{F}^{2}/\mathbb{E}[\|\E_{d}\|_{F}^{2}]=\|\M_{d}\|_{F}^{2}/(\sigma_{d}^{2}np_{d})$. We vary the number of views ($D=2$ or $D=3$), and consider two cases for each $D$: (i)~orthogonal (all joint, partially-shared and individual structures are orthogonal); (ii) non-orthogonal (some structures are non-orthogonal). Appendix \ref{supp:sim} provides additional data generation details. Given an estimated $\widehat{\M}=[\widehat{\M}_{1}\ \dots \widehat{\M}_{D}]$, we compare all methods in terms of the estimated ranks and the scaled squared Frobenius norm error defined by
\begin{equation}\label{sfb}
L_{F}(\M,\widehat{\M}) = \sum_{d=1}^{D}\frac{1}{\|\M_{d}\|_{F}^{2}}\|\M_{d}-\widehat{\M}_{d}\|_{F}^{2}.
\end{equation}

\begin{figure}[H]
    \centering
    \includegraphics[scale = 0.7]{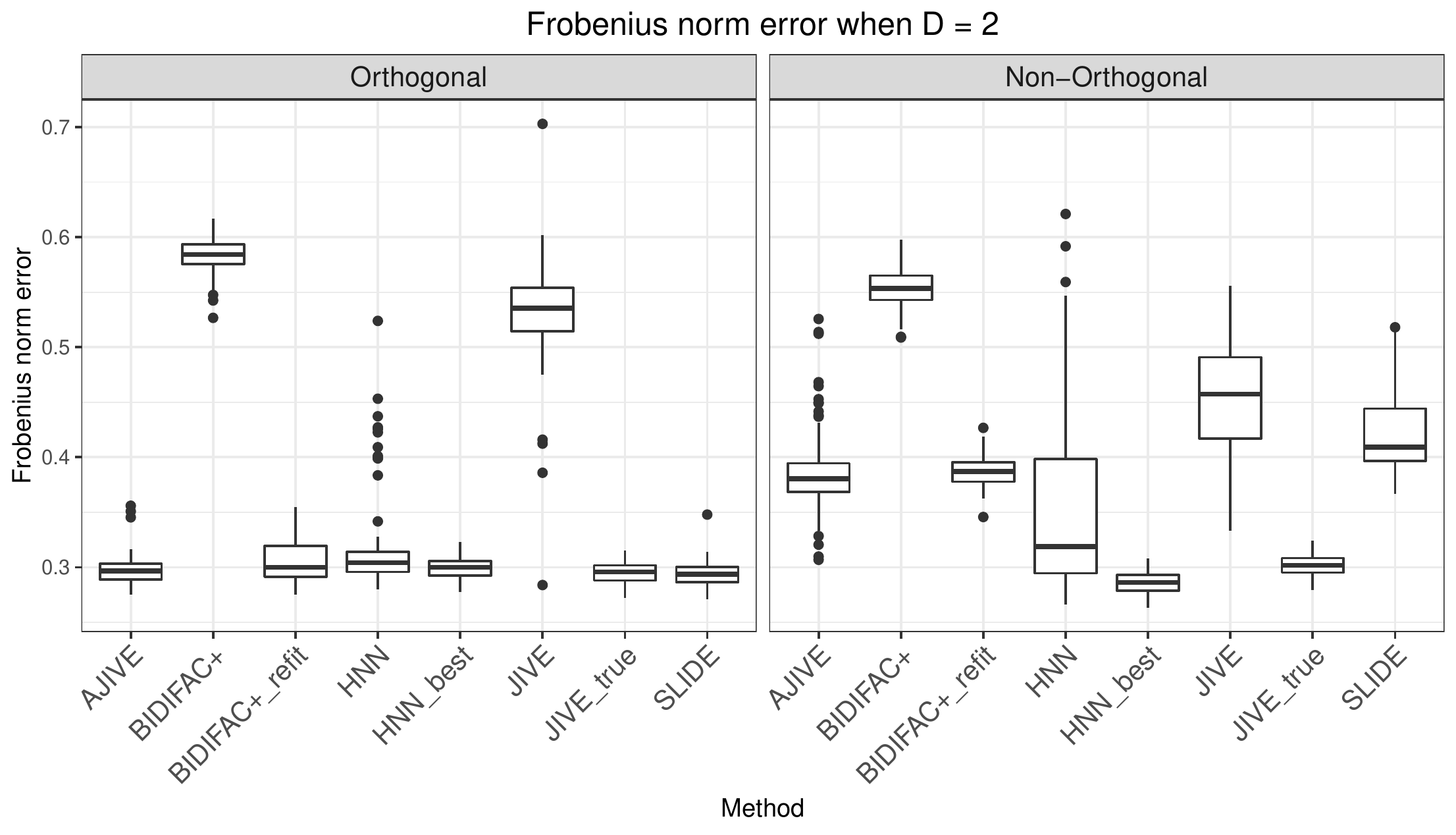}         
    \includegraphics[scale = 0.7]{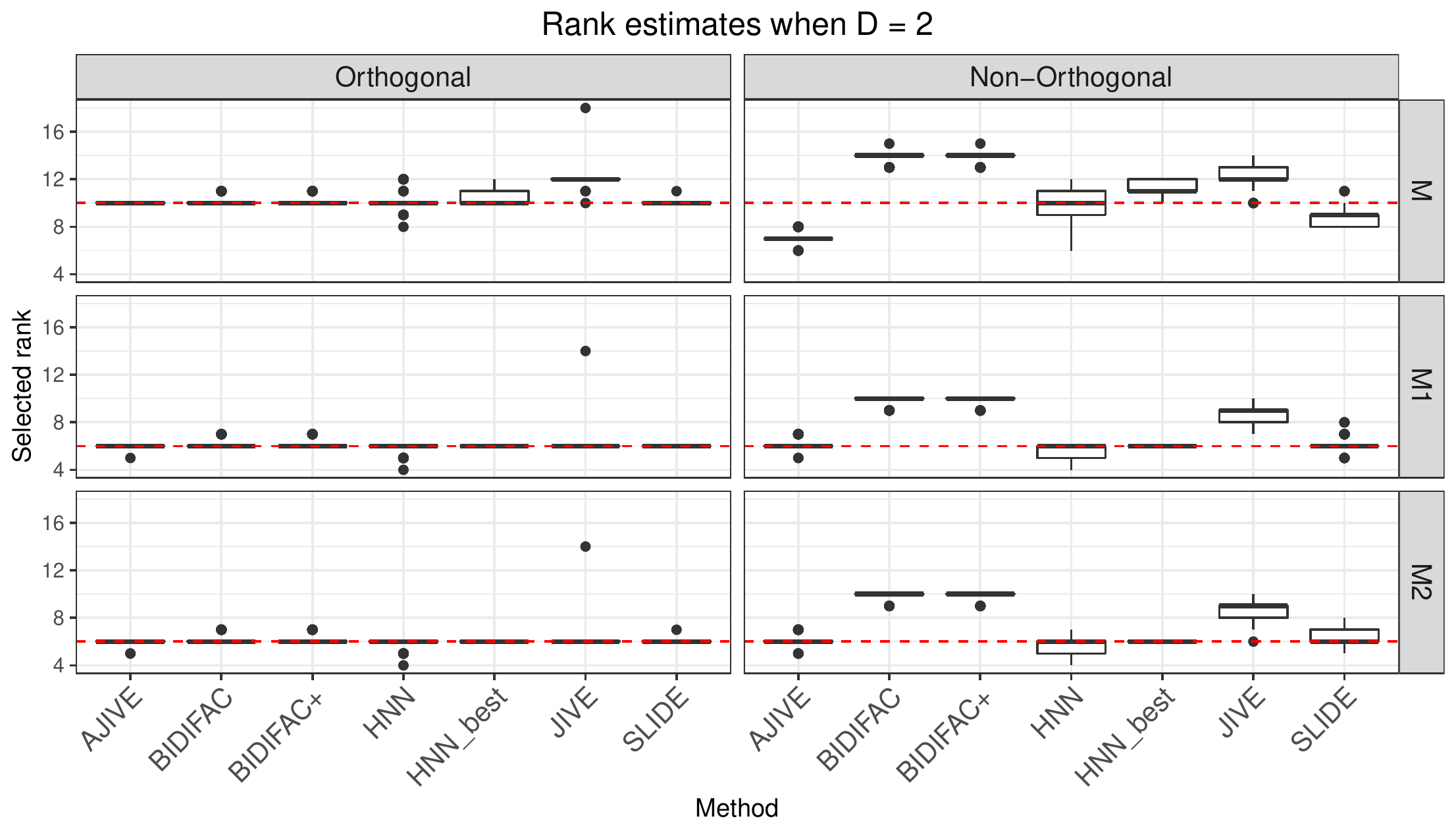}     
    \caption{Boxplots of scaled squared Frobenius norm errors \eqref{sfb} and rank estimates of the concatenated signals and each $\M_{d}$ over 100 independent replication for Orthogonal scheme (left column) and Non-orthogonal scheme (right column) when $D = 2$. The true ranks are indicated by the red dotted line}\label{sim:fig_d2}
\end{figure}

Figures \ref{sim:fig_d2} shows the results for $D=2$ setting over 100 replications. In the orthogonal case, AJIVE and SLIDE are the best performing methods, followed closely by HNN and BIDIFAC+\_refit. Good performance of SLIDE is expected in the orthogonal case as the identifiability conditions are satisfied. Good performance of AJIVE can be attributed to its accurate ranks estimation, as no partially-shared signals exist when $D=2$. JIVE and BIDIFAC+ perform the worst.  
Comparing the ranks reveals that JIVE overestimates the rank of $\bM$ while correctly estimating the ranks of individual $\bM_d$, hence it under-estimates the rank of joint structure. Since JIVE with true ranks (JIVE\_true) performs as well as AJIVE and SLIDE, the results indicate that JIVE's poor performance in this setting is due to rank mis-identification. As expected, BIDIFAC+ performs significantly worse than BIDIFAC+\_refit due to the bias associated with the penalization in original BIDIFAC+ solution. In the non-orthogonal case, the performance of both AJIVE and SLIDE deteriorates, with HNN becoming the best performing method. The deterioration of SLIDE is expected as non-orthogonal individual structures violate SLIDE's identifiability conditions \citep{gaynanova2019structural}. The deterioration of AJIVE could be attributed to rank mis-identification, as it significantly under-estimates the total rank of $\bM$. The median estimated ranks of HNN always match the true ranks (in both orthogonal and non-orthogonal cases).

\begin{figure}[H]
    \centering
    \includegraphics[scale = 0.73]{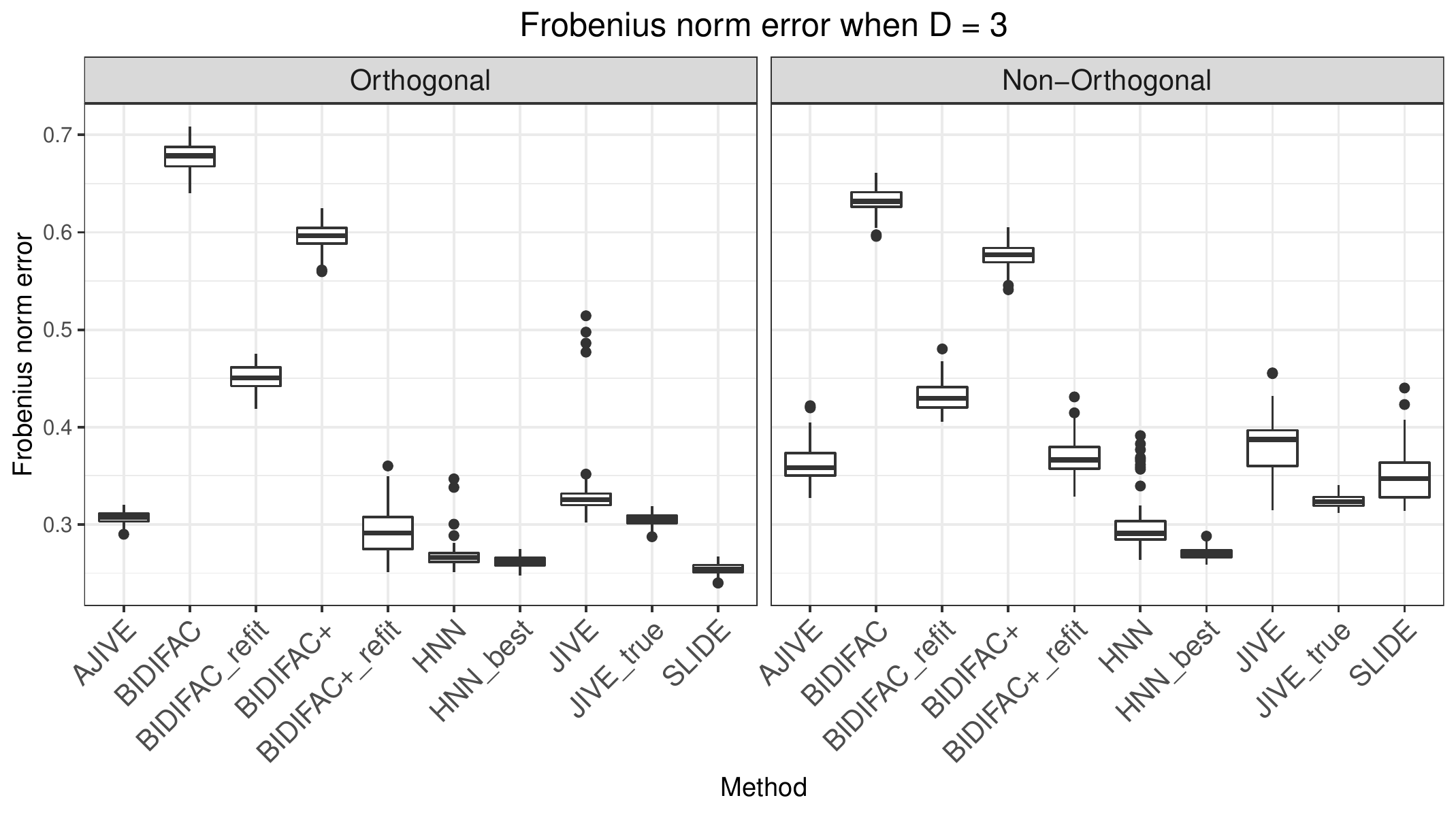}         
    \includegraphics[scale = 0.73]{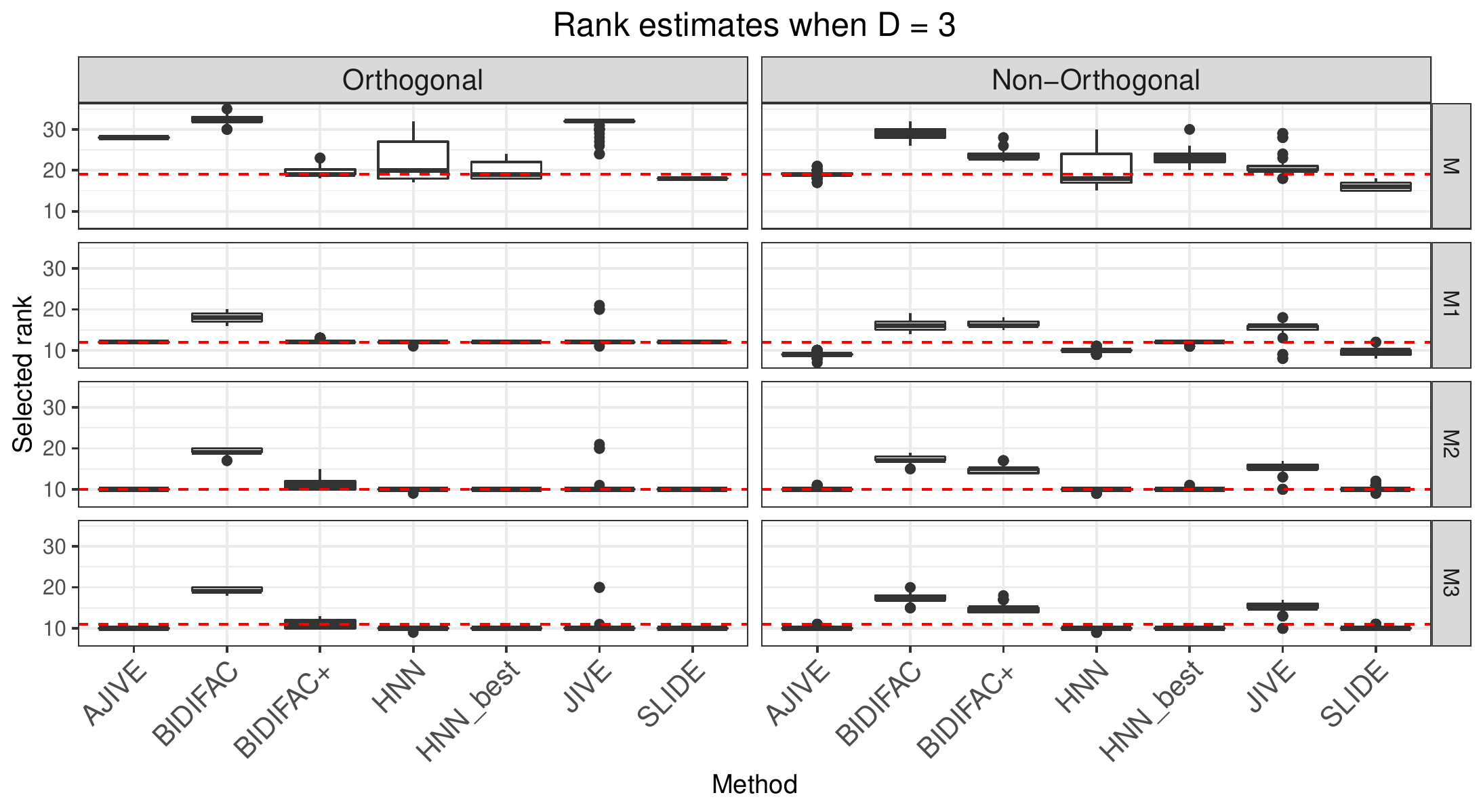}         
    \caption{Boxplots of scaled squared Frobenius norm errors \eqref{sfb} and rank estimates of the concatenated signals and each $\M_{d}$ over 100 independent replication for Orthogonal scheme (left column) and Non-orthogonal scheme (right column) when $D = 3$. The true ranks are indicated by the red dotted line}\label{sim:fig_d3}    
\end{figure}

Figure~\ref{sim:fig_d3} shows the results for $D=3$ setting over 100 replications. Unlike $D=2$ setting, HNN performs better than AJIVE in both orthogonal and non-orthogonal cases, and is only second to SLIDE in the orthogonal case (albeit close in both errors and estimated ranks). AJIVE's worse performance is expected since it does not account for partially-shared structures. JIVE performs better or similar to AJIVE, but worse than HNN, SLIDE and BIDIFAC+\_refit. As expected, the refitted versions of BIDIFAC are better than the original ones, with BIDIFAC+ performing better than BIDIFAC. As in $D=2$ case, SLIDE's performance is significantly worse in non-orthogonal case due to identifiability issues. In non-orthogonal case, the median estimated ranks for HNN are the closest to true ranks.

Surprisingly, BIDIFAC and BIDIFAC+ have significantly worse performance than other methods in our experiments. This is due to the bias caused by the penalization as the resulting singular values from both BIDIFAC and BIDIFAC+ are shrunken too much relative to the true values. In Figures \ref{sim:fig_d2}-\ref{sim:fig_d3}, both BIDIFAC\_refit and BIDIFAC+\_refit show much smaller Frobenius norm errors compared to the corresponding results without refitting.

Overall, the proposed HNN performs the best in terms of median estimation errors and median estimated ranks. As expected, the errors of oracle HNN\_best are always better. However, the difference between the two medians is negligible compared to the difference with other methods. This supports the good performance of the proposed tuning parameter selection procedure. At the same time, HNN shows high variability across replications, especially in the non-orthogonal two-view case. As shown in Appendix \ref{supp:sim}, this extra variability is due to the randomness in BCV splits. An occasional ``poor" split may lead to rank under-estimation, subsequently giving higher signal estimation errors, whereas a different split on the same data has no such issue. In practice, we recommend running the BCV procedure with several split choices to assess the ranks that are selected most consistently across splits.
In the orthogonal case, SLIDE has an advantage over HNN as it has similar or slightly higher accuracy, and lower variance. However, SLIDE performs significantly worse in non-orthogonal cases, which is further supported by additional results in Appendix \ref{supp:simres}. Since in practice it is difficult to assess a priori whether orthogonality condition holds, HNN has an advantage over SLIDE.

\section{Analysis of multi-tissue gene expression data}\label{sec:real}

The p53 gene plays an important role in cell regulation with impact on aging processes and cancer \citep{seim2016gene,tanikawa2017transcriptional}. However, the tissue-specific nature of p53 gene expressions makes it difficult to develop targeted therapies \citep{vlatkovic2011tissue}. Thus, it is crucial to identify patterns in p53 gene expression that are shared and unique across multiple tissues. To identify such patterns, we analyze gene expression data corresponding to p53 signaling pathway from three tissues (muscle, blood and skin). The original data are available from the Genotype-Tissue Expression project \citep{gtex2020gtex}, we use the preprocessed data by \cite{li2017incorporating} that is available at \url{https://github.com/reagan0323/SIFA}. The resulting data contain measurements on 191 genes from 204 matched samples corresponding to three tissues: $\X_{1}\in\mathbb{R}^{204\times191}$ for muscle, $\X_{2}\in\mathbb{R}^{204\times191}$ for blood and $\X_{3}\in\mathbb{R}^{204\times191}$ for skin. Our goal is to identify shared, partially-shared and individual signal structures across three tissues, and compare the results across different methods. 

In the first three rows of Table \ref{real:tab1}, we compare HNN with JIVE, AJIVE, SLIDE, BIDIFAC and BIDIFAC+ on each view in terms of (i) the total estimated rank; (ii) the percentage of explained variation (the ratio of the squared Frobenius norm of the estimated signal to the squared Frobenius norm of data). JIVE, AJIVE and BIDIFAC estimate smaller ranks compared to SLIDE, BIDIFAC+ and HNN. Furthermore, HNN has the highest total ranks for each view, with highest percentage of variance explained. While larger explained variance can be partially attributed to larger ranks, this is not the case for muscle and skin tissues. For muscle, HNN has higher percentage of variance explained  than SLIDE and BIDIFAC+\_refit with the same rank of 16. For skin, HNN has higher percentage than SLIDE with the same rank of 19. To provide a more fair comparison, we adjust HNN estimates to match the ranks of other methods by applying the singular value decomposition to each $\widehat \bM_d$ and only keeping the top singular vectors (matching the rank). The adjusted HNN estimates still explain more variation in each view than other methods (Figure \ref{fig:svd_GTEx} in Appendix \ref{supp:real}). Overall, these comparisons suggest that HNN provides a better fit to the data. 

\begin{table}[H]
    \centering
    \scalebox{0.78}{
    \begin{tabular}{|c|c|c|c|c|c|c|} 
    \hline
     & JIVE & AJIVE & SLIDE & HNN & BIDIFAC w/wo refit & BIDIFAC+ w/wo refit \\
    \hline
    Muscle  & 13 (66.3\%) & 12 (66.9\%) & 16 (69.2\%) & 16 (71.4\%) & 13 (64.4\%/18.8\%) & 16 (65.6\%/19.0\%) \\ 
     Blood  & 11 (72.4\%) & 5 (68.0\%) & 12 (77.7\%) & 17 (82.7\%) & 11 (75.3\%/30.3\%) & 12 (75.7\%/30.2\%) \\
    Skin & 15 (61.1\%) & 11 (66.4\%) & 19 (73.3\%) & 19 (75.7\%) & 14 (67.4\%/16.4\%) & 17 (68.4\%/16.6\%) \\
    \hline
    Joint & 2 & 1 & 3 & 1 & 5 & 2 \\
    Muscle \& Blood & NA & NA & 0 & 3 & NA & 2\\
    Muscle \& Skin & NA & NA & 0 & 2 & NA & 5\\
    Blood \& Skin & NA & NA & 1 & 2 & NA & 2\\
    Muscle only & 11 & 11 & 13 & 10 & 8 & 7\\
    Blood only & 9 & 4 & 8 & 11 & 6 & 6\\
    Skin only & 13 & 10 & 15 & 14 & 9 & 8 \\
    \hline
    \end{tabular}
    }
    \caption{Table of the rank estimates of each view with the percentage of explained variation (Rows 1-3) and the estimated ranks of each signal structures (Rows 4-10). For both BIDIFAC and BIDIFAC+, we present the corresponding results with and without refitting.}
    \label{real:tab1}
\end{table}

The last seven rows in Table \ref{real:tab1} show the ranks of the corresponding structures identified by each method. For all methods, individual structures have higher ranks than shared structures, which is consistent with known tissue-specificity of  p53 gene expressions  \citep{seim2016gene,tanikawa2017transcriptional}. HNN identifies more partially-shared structures compared to SLIDE but less than BIDIFAC+. Since JIVE, AJIVE and BIDIFAC are unable to identify partially-shared structures directly, we also apply these methods separately to each pair of views. If the joint rank estimated on a pair of views is larger than the joint rank estimated on all three views, there is evidence for partially-shared structures. Table \ref{real:tab2} shows estimated ranks for JIVE, AJIVE and BIDIFAC when applied to each pair of views. For Muscle \& Skin, the joint ranks (3 for JIVE, 3 for AJIVE and 6 for BIDIFAC) are larger than the corresponding joint ranks for all three views (2 for JIVE, 1 for AJIVE and 5 for BIDIFAC), suggesting a missing partially-shared structure between the muscle and skin tissues. From Table \ref{real:tab1}, BIDIFAC$+$ and HNN are the only methods that identify partial-sharing between Muscle \& Skin.  To compare how partially-shared structures from HNN relate to other methods, in Figure \ref{fig:angles2} in Appendix \ref{supp:real} 
we illustrate cosines of principal angles between HNN partially-shared structures and the corresponding individual structures from the other approaches. For both Muscle \& Blood and Muscle \& Skin, we find that rank one subspace of corresponding partially-shared structure from HNN is identified as individual for Muscle by JIVE and AJIVE (cosine value close to one). For Blood \& Skin, the partially-shared structures of HNN are distinct, and are not identified as individual by any of the other methods. 

\begin{table}[H]
    \centering
    \scalebox{0.9}{
    \begin{tabular}{|c|c|c|c|c|c|c|c|c|c|} 
    \hline
    & \multicolumn{3}{c|}{JIVE} & \multicolumn{3}{c|}{AJIVE} & \multicolumn{3}{c|}{BIDIFAC} \\
    \cline{2-10}
    & Joint & Indiv 1 & Indiv 2 & Joint & Indiv 1 & Indiv 2 & Joint & Indiv 1 & Indiv 2 \\  
    \hline
    Muscle\ \&\ Blood & 2 & 11 & 9 & 1 & 11 & 4 & 4 & 8 & 6 \\
    Muscle\ \&\ Skin & 3 & 11 & 15  & 3 & 9 & 8 & 6 & 7 & 8 \\ 
    Blood\ \&\ Skin & 2 & 8 & 13 & 1 & 4  &  10 & 3 & 6 & 9 \\ 
    \hline
    \end{tabular}
    }
    \caption{Table of the estimated ranks of the structures from JIVE, AJIVE and BIDIFAC with each pairs of the GTEx data}
    \label{real:tab2}
\end{table}

The numerical results in Section~\ref{sec:sim} coupled with comparisons of HNN with other methods in terms of variance explained and identified ranks suggest that HNN provides new insights into specificity of p53 pathway gene expressions across tissues. The presence of partially-shared structures across each pair of tissues supports the existing research that each pair has unique similarity depending on the underlying characteristics used for comparison. \citet{sonawane2017understanding} clustered tissues into communities based on tissue-specific targeting patterns: both Skin and Blood tissues belong to the cell proliferation community, whereas Muscle tissue belongs to a distinct extracellular structure community. 
\citet{gtex2020gtex} considered pairwise associations between tissues based on cis-eQTLs, expression quantitative trait loci located near the gene of origin. They found that Skin and Muscle have a high pairwise association, whereas the cis-eQTL-based association of each tissue with Blood was considerably lower. 
\citet{oliva2020impact} considered the impact of gender on tissue-expressions, with considerably higher number of sex-biased genes in Skin tissue compared to Muscle and Blood tissues. Consequently, they found that Muscle and Blood tissues co-cluster based on the effect size of sex-biased genes, with Skin tissue being significantly more distinct. While these prior analyses were not restricted to p53 signaling pathway, it is known that p53 pathway plays a crucial role in cell proliferation \citep{seim2016gene} and is enriched for sex-biased genes \citep{lopes2020genome}. The identified shared, partially-shared and individual signal by HNN provide a principled data-driven foundation for further investigation of distinct biological characteristics affecting p53 expression. 

\section{Conclusion}\label{sec:conclude}
In this work, we formally introduce a model for joint, partially-shared, and individual signal structures in multi-view data based on hierarchical levels. We also propose a new hierarchical nuclear norm penalization approach for signal estimation and a simple refitting step for bias adjustment. Our simulation studies indicate that our method outperforms competing methods in terms of signal recovery and rank estimation, particularly when not all signal structures are orthogonal.

While the proposed BCV-based tuning parameter selection scheme performs well in our numerical studies, a large number of tuning parameters still needs to be chosen,  leading to high computational burden. It would be of interest to explore a way of choosing tuning parameters that is analogous to the SURE approach in single-view nuclear norm penalization. This approach, however, is quite challenging because Algorithm \ref{algo:dbfb} does not admit a closed-form solution. Theoretical analysis of our method is also an interesting future study, which would be possible thanks to considerable development of theories in single-view nuclear norm penalization \citep{bach2008consistency,negahban2011estimation}.

\section*{Acknowledgements}
This research was supported by the National Science Foundation grant CCF-1934904. IG was partially supported by NSF DMS CAREER-2044823.

\appendix
\section{Supplementary material}

Appendix \ref{supp:method} and Appendix \ref{supp:sim} provide supplementary materials for Section \ref{sec:method} and Section \ref{sec:sim} in the main paper, respectively. Appendix \ref{supp:simres} contains additional simulation studies. Appendix \ref{supp:real} provides additional analysis of GTEx data.

\subsection{Supplementary material for Section \ref{sec:method}}\label{supp:method}

\subsubsection{Example with two views}\label{supp:jive}

\begin{example}\label{ex:d2}
\normalfont
Consider $D=2$ with the signal matrices:
\[
\M_{1} =\begin{bmatrix}
1 & 0 \\
0 & 1 \\
0 & 0 \\
\end{bmatrix},\quad \M_{2} = \begin{bmatrix}
1 & 0 \\
0 & 1\\
0 & 1 \\
\end{bmatrix}.    
\]
JIVE \citep{lock2013joint}, COBE \citep{zhou2016group} and AJIVE \citep{feng2018angle} define the joint structure as the intersection of the column spaces: 
\[
\mathcal{C}(\M_{1})\cap\mathcal{C}(\M_{2}) = \text{span}\left\{ [1\quad 0\quad 0]^\top\right\},
\]
which represents a signal structure shared by both views. The individual structures are defined as the column space of the remaining non-intersecting signal, that is the projection of the signal onto the orthogonal complement of the joint structure  $\mathcal{C}(\M_{1})\cap\mathcal{C}(\M_{2})$: 
\[
\begin{aligned}
&\text{Individual view 1:}\quad \mathcal{C}(\P_{\mathcal{C}(\M_{1})\cap\mathcal{C}(\M_{2})}^\perp\M_{1}) = \text{span}\left\{ [0\quad 1\quad 0]^\top\right\},\\ 
&\text{Individual view 2:}\quad \mathcal{C}(\P_{\mathcal{C}(\M_{1})\cap\mathcal{C}(\M_{2})}^\perp\M_{2}) = \text{span}\left\{ [0\quad 1\quad 1]^\top \right\}.
\end{aligned}    
\]
The individual structures represent view-specific patterns. By construction, they are orthogonal to the joint structure, and have zero intersection with each other (albeit not necessarily orthogonal to each other).
\end{example}

\subsection{SLIDE identifiability example}

Web Appendix A.3 in \cite{gaynanova2019structural} provides SLIDE decomposition given $\M_{d}$ when $D=3$. 
Starting with $\R_{d}^{\text{SLIDE}}=\M_{d}$, we identify each structure as follows: 

\begin{enumerate}

    \item (Individual structure) For view $d=1,2,3$: 
    
        \begin{enumerate}
    
            \item Set $\U^{(d)}$, the individual score for view $d$, to be the orthonormal basis vectors for $\mathcal{C}\left(\P_{\mathcal{C}(\R_{k}^{\text{SLIDE}})\cup\mathcal{C}(\R_{l}^{\text{SLIDE}})}^\perp    \R_{d}^{\text{SLIDE}}\right)$ for distinct $d,k,l$. 
            
            \item Set $\V^{(d)}=\left(\P_{\mathcal{C}(\R_{k}^{\text{SLIDE}})\cup\mathcal{C}(\R_{l}^{\text{SLIDE}})}^\perp\R_{d}^{\text{SLIDE}}\right)^\top\U^{(d)}$ to be the loading matrix corresponding $\U^{(d)}$.
            
            \item Update the residual $\R_{d}^{\text{SLIDE}}=(\I - \P_{\mathcal{C}(\R_{k}^{\text{SLIDE}})\cup\mathcal{C}(\R_{l}^{\text{SLIDE}})}^\perp)\R_{d}^{\text{SLIDE}}$.
            
        \end{enumerate}
    
    \item (Partially-shared structure) For $(k,l) = (1,2),(1,3),(2,3)$: 
    
        \begin{enumerate}
    
            \item Set $\U^{(kl)}$, the individual score for view $d$, to be the orthonormal basis vectors for $\mathcal{C}\left(\P_{\mathcal{C}(\R_{d}^{\text{SLIDE}})}^\perp [\R_{k}^{\text{SLIDE}}\ \R_{l}^{\text{SLIDE}}]\right)$ for $d\neq k$ and $d\neq l$.
            
            \item Set $\V^{(kl)}=\left(\P_{\mathcal{C}(\R_{d}^{\text{SLIDE}})}^\perp[\R_{k}^{\text{SLIDE}}\ \R_{l}^{\text{SLIDE}}]\right)^\top\U^{(kl)}$ to be the loading matrix corresponding $\U^{(kl)}$.
            
            \item Update the residual $[\R_{k}^{\text{SLIDE}}\ \R_{l}^{\text{SLIDE}}]=(\I - \P_{\mathcal{C}(\R_{d}^{\text{SLIDE}})}^\perp)[\R_{k}^{\text{SLIDE}}\ \R_{l}^{\text{SLIDE}}]$.
            
        \end{enumerate}    
    
    \item (Joint structure) 
    
        \begin{enumerate}
        
            \item Set the joint score $\U^{(123)}$ to be the orthonormal basis vectors for $\mathcal{C}(\R^{\text{SLIDE}})$ where $\R^{\text{SLIDE}}=[\R_{1}^{\text{SLIDE}}\  \R_{2}^{\text{SLIDE}}\  \R_{3}^{\text{SLIDE}}]$.
            
            \item Set $\V^{(123)}=[ \R_{1}^{\text{SLIDE}}\  \R_{2}^{\text{SLIDE}}\  \R_{3}^{\text{SLIDE}}]^\top \U^{(123)}$.
            
        \end{enumerate}
    
\end{enumerate}

\begin{example}\label{ex:slide}
\normalfont

Recall that the signal matrices in Example \ref{ex:d3} are
\begin{equation}\label{ex:signals}
\begin{aligned}
&\M_{1} =\begin{bmatrix}
1 & 1 & 1 \\
-1 & 1 & 0 \\
1 & 0 & -1 \\
-1 & 0 & 0 \\
1 & 0 & 0 \\
\end{bmatrix},\quad \M_{2} =\begin{bmatrix}
1 & 1 & 0 \\
-1 & 1 & 0 \\
1 & 0 & 1 \\
-1 & 0 & 0 \\
1 & 0 & -1 \\
\end{bmatrix}, \quad 
\M_{3} = \begin{bmatrix}
1 & 1 & 0 \\
-1 & 0 & -1 \\
1 & -1 & 0 \\
-1 & 0 & 2 \\
1 & 0 & 1 \\
\end{bmatrix}.    
\end{aligned}
\end{equation}

However, the signal matrices in \eqref{ex:signals} do not admit a unique SLIDE decomposition because its identifiability condition (the orthogonality between each signal structure) is violated. 
To see this, we apply the above construction process of SLIDE decomposition to $\M_{d}$'s in \eqref{ex:signals} according to two different orders for score and loading calculation:

\begin{itemize}

    \item Order 1: We calculate the individual scores and loadings of 1st, 2nd and 3rd view. And the partially-shared scores and loadings of (i) 1st and 2nd views; (ii) 1st and 3rd views; (iii) 2nd and 3rd views are calculated. This order gives the following decomposition of the signal matrices in \eqref{ex:signals}:
    \begin{equation}\label{decomp1}
    [\M_{1}\ \M_{2}\ \M_{3}] = \U^{(123)}(\V^{(123)})^\top + [\U^{(12)}(\V^{(12)})^\top\ \mathbf{0}] + [\mathbf{0}\ \U^{(2)}(\V^{(2)})^\top\ \mathbf{0}], 
    \end{equation}
    where the score and loading matrices after rounding to the second decimal are 
    \[
    \U^{(123)}=\begin{bmatrix}
    -0.45 & 0.18 & -0.68 \\
    0.45 & 0.39 & 0.11 \\
    -0.45 & -0.18 & 0.68 \\
    0.45 & -0.79 & -0.21 \\
    -0.45 & -0.39 & -0.11 \\
    \end{bmatrix},\quad  \V^{(123)}=\begin{bmatrix}
    -2.24 & 0 & 0 \\
    0 & 0.58 & -0.58 \\
    0 & 0.37 & -1.37 \\
    -2.24 & 0 & 0 \\
    0 & 0.58 & -0.58 \\
    0 & 0.21 & 0.79 \\
    -2.24 & 0 & 0 \\
    0 & 0.37 & -1.37 \\
    0 & -2.37 & -0.63 \\    
    \end{bmatrix},
    \]
    \[
    \U^{(12)}=\begin{bmatrix}
    -0.43 \\
    -0.72 \\
    -0.43 \\
    -0.29 \\
    -0.14 \\
    \end{bmatrix},\quad  \V^{(12)}=\begin{bmatrix}
    0 \\
    -1.15 \\
    0 \\
    0 \\
    -1.15 \\
    -0.29 \\
    \end{bmatrix},\quad
    \U^{(2)}=\begin{bmatrix}
    -0.34 \\
    0.34 \\
    -0.34 \\
    -0.22 \\
    0.78 \\
    \end{bmatrix}, \quad \V^{(2)}=\begin{bmatrix}
    0 \\
    0 \\
    -1.12 \\
    \end{bmatrix}.
    \]

    \item Order 2:  We calculate the individual scores and loadings of 3rd, 1st and 2nd view. And the partially-shared scores and loadings of (i) 2nd and 3rd views; (ii) 1st and 3rd views; (iii) 1st and 2nd views are calculated. This gives the following decomposition of the signal matrices in \eqref{ex:signals}:
    \begin{equation}\label{decomp2}
    [\M_{1}\ \M_{2}\ \M_{3}] = \U^{(123)}(\V^{(123)})^\top + [\mathbf{0}\ \U^{(23)}(\V^{(23)})^\top] + [\mathbf{0}\ \mathbf{0}\ \U^{(3)}(\V^{(3)})^\top] 
    \end{equation}
    where the score and loading matrices after rounding to the second decimal are 
    \[
    \begin{aligned}
    \U^{(123)}&=\begin{bmatrix}
    -0.45 & -0.82 & 0 \\
    0.45 & -0.41 & -0.71 \\
    -0.45 & 0.41 & -0.71 \\
    0.45 & 0 & 0 \\
    -0.45 & 0 & 0 \\
    \end{bmatrix},\quad  \V^{(123)}=\begin{bmatrix}
    -2.24 & 0 & 0 \\
    0 & -1.22 & -0.71 \\
    0 & -1.22 & 0.71 \\
    -2.24 & 0 & 0 \\
    0 & -1.22 & -0.71 \\
    0 & 0.41 & -0.71 \\
    -2.24 & 0 & 0 \\
    0 & -1.22 & 0.71 \\
    0 & 0.41 & 0.71 \\    
    \end{bmatrix}, \\    
    \U^{(23)}&=\begin{bmatrix}
    -0.29 \\
    0.29 \\
    -0.29 \\
    0 \\
    0.87 \\
    \end{bmatrix},\quad  \V^{(23)}=\begin{bmatrix}
    0 \\
    0 \\
    -1.15 \\
    0 \\
    0 \\
    0.58 \\
    \end{bmatrix},\quad
    \U^{(3)}=\begin{bmatrix}
    0.22 \\
    -0.22 \\
    0.22 \\
    0.89 \\
    0.22 \\
    \end{bmatrix}, \quad \V^{(3)}=\begin{bmatrix}
    0 \\
    0 \\
    2.24 \\
    \end{bmatrix}.    
    \end{aligned}
    \]

\end{itemize}

While the decomposition \eqref{decomp1} identifies (a) the joint structure with rank 3; (b) the partially-shared structure of 1st and 2nd views with rank 1; (c) the individual structure of 2nd view with rank 1, the decomposition \eqref{decomp2} identifies (a) the joint structure with rank 3; (b) the partially-shared structure of 2nd and 3rd views with rank 1; (c) the individual structure of 3rd view with rank 1. 

\end{example}

\subsection{Derivation of Algorithm \ref{algo:dbfb}}

Given $\tilde{x}\in\mathbb{R}^{N\times1}$, \cite{abboud2017dual} consider the following optimization problem:
\begin{equation}\label{dbfb:opt}
    \minimize_{x\in\mathbb{R}^{N}} \sum_{j=1}^{J} h_{j}(\mathbf{A}_j x) + \frac{1}{2}\|x-\tilde{x}\|^{2}
\end{equation}
where $\mathbf{A}_{j}\in\mathbb{R}^{m_{j} \times N}$ and $h_{j}$ is a convex function from $\mathbb{R}^{m_{j}}$ to $\mathbb{R}$. \cite{abboud2017dual} present Algorithm \ref{supp:algo} to solve \eqref{dbfb:opt}. Before deriving Algorithm \ref{algo:dbfb}, we first describe how the notations in Algorithm \ref{supp:algo} and \eqref{dbfb:opt} can be translated into ours in Table \ref{deriv:tab1}. 


\begin{algorithm}[!t]
\caption{Algorithm (40) in \cite{abboud2017dual}}\label{supp:algo}
\begin{algorithmic}
\STATE \text{Input}: $\gamma\in(0,2)$,  $y_{0}^{j}\in\mathbb{R}^{m_{j}}$ and $\B_{j}$ such that $\B_{j}\succ \mathbf{0}_{j}$ with $\B_{j}\succeq\A_{j}\A_{j}^\top$ (i.e., $\B_{j}-\A_{j}\A_{j}^\top$ is positive semi-definite) for $1\leq j \leq J$. 
\FOR{$t=0,1,\dots$}
\FOR{$j=1,\dots, J$}
\STATE  $\tilde{y}_{t}^{j} = y_{t}^{j}+\gamma\B_{j}^{-1}\A_{j}x_{t}$ 
\STATE $y_{t+1}^{j} = \tilde{y}_{t}^{j}-\gamma\B_{j}^{-1}\text{prox}_{\gamma\B_{j}^{-1},h_{j}}\left(\gamma^{-1}\B_{j}^{-1}\tilde{y}_{t}^{j} \right)$
\STATE where $\text{prox}_{\B, \psi}(\tilde{x})$ is the proximity operator of $\psi$ at $\tilde{x}$ relative to the metric induced by 
\STATE $\B$ such that $\text{prox}_{\B, \psi}(\tilde{x})=\argmin_{x\in\mathbb{R}^{N}} \psi(x)+ \frac{1}{2}\|x-\tilde{x}\|_{B}^{2}$ for $\|x\|_{B}^{2}=x^\top B x$ 
\ENDFOR
\STATE  $x_{t+1} = x_{t} - \sum_{j=1}^{J}\A_{j}^\top\left( y_{t+1}^{j} - y_{t}^{j}\right)$
\ENDFOR
\end{algorithmic}
\end{algorithm}

For three views, recall that the optimization problem with the hierarchical nuclear norm penalty is
\begin{equation}\label{supp:opt:3views}
\minimize_{\M_{1},\M_{2},\M_{3}}  \biggl\{ \frac{1}{2}\sum_{d=1}^{3}\left\|\X_{d}-\M_{d}\right\|_{F}^{2} + \sum_{d=1}^{3}\lambda_{d}\|\M_{d}\|_{*}+\sum_{k,l=1,k<l}^{3}\lambda_{kl}\|\M_{kl}\|_{*}  +\lambda\|\M\|_{*} \biggr\},    
\end{equation}
where $\M_{kl}=[\M_{k}\ \M_{l}]\in\mathbb{R}^{n \times (p_{k}+p_{l})}$. It is straightforward to verify that \eqref{dbfb:opt} reduces to \eqref{supp:opt:3views} after plugging the corresponding notations in Table \ref{deriv:tab1}. In the following, we similarly check how the updates in Algorithm \ref{algo:dbfb} of the main paper correspond to the three update formulae in Algorithm \ref{supp:algo}.

\begin{enumerate}
    \item Plugging the corresponding notation in the second column of Table \ref{deriv:tab1} into $\tilde{y}_{t}^{j} = y_{t}^{j}+\gamma\B_{j}^{-1}\A_{j}x_{t}$ and $y_{t+1}^{j} = \tilde{y}_{t}^{j}-\gamma\B_{j}^{-1}\text{prox}_{\gamma\B_{j}^{-1},h_{j}}\left(\gamma^{-1}\B_{j}^{-1}\tilde{y}_{t}^{j} \right)$ from Algorithm \ref{supp:algo}, we have
    \begin{equation}\label{upd:duals}
    \begin{aligned}
    \text{vec}(\D_{1}^{(t+1)}) &= \text{vec}(\D_{1}^{(t)}+\gamma\M_{1}^{(t)})-\gamma \text{prox}_{\gamma\mathbf{I}_{np_{1}}, h_{1}}(\gamma^{-1}\text{vec}(\D_{1}^{(t)}+\gamma\M_{1}^{(t)})), \\ 
    \text{vec}(\D_{2}^{(t+1)}) &= \text{vec}(\D_{2}^{(t)}+\gamma\M_{2}^{(t)})-\gamma \text{prox}_{\gamma\mathbf{I}_{np_{2}}, h_{2}}(\gamma^{-1}\text{vec}(\D_{2}^{(t)}+\gamma\M_{2}^{(t)})), \\ 
    \text{vec}(\D_{3}^{(t+1)}) &= \text{vec}(\D_{3}^{(t)}+\gamma\M_{3}^{(t)})-\gamma \text{prox}_{\gamma\mathbf{I}_{np_{3}}, h_{3}}(\gamma^{-1}\text{vec}(\D_{3}^{(t)}+\gamma\M_{3}^{(t)})), \\     
    \text{vec}(\D_{12}^{(t+1)}) &= \text{vec}(\D_{12}^{(t)}+\gamma\M_{12}^{(t)})-\gamma \text{prox}_{\gamma\mathbf{I}_{n(p_{1}+p_{2})}, h_{4}}(\gamma^{-1}\text{vec}(\D_{12}^{(t)}+\gamma\M_{12}^{(t)})), \\     
    \text{vec}(\D_{13}^{(t+1)}) &= \text{vec}(\D_{13}^{(t)}+\gamma\M_{13}^{(t)})-\gamma \text{prox}_{\gamma\mathbf{I}_{n(p_{1}+p_{3})}, h_{5}}(\gamma^{-1}\text{vec}(\D_{13}^{(t)}+\gamma\M_{13}^{(t)})), \\     
    \text{vec}(\D_{23}^{(t+1)}) &= \text{vec}(\D_{23}^{(t)}+\gamma\M_{23}^{(t)})-\gamma \text{prox}_{\gamma\mathbf{I}_{n(p_{2}+p_{3})}, h_{6}}(\gamma^{-1}\text{vec}(\D_{23}^{(t)}+\gamma\M_{23}^{(t)})), \\     
    \text{vec}(\D^{(t+1)}) &= \text{vec}(\D^{(t)}+\gamma\M^{(t)})-\gamma \text{prox}_{\gamma\mathbf{I}_{n(p_{1}+p_{2}+p_{3})}, h_{7}}(\gamma^{-1}\text{vec}(\D^{(t)}+\gamma\M^{(t)})). \\        
    \end{aligned}        
    \end{equation}
    By its definition in Algorithm \ref{supp:algo}, each proximity operator in \eqref{upd:duals} can be further simplified. Letting $\widetilde{\D}_{1}^{(t)}=\D_{1}^{(t)}+\gamma\M_{1}^{(t)}$, for example, we have 
    \[
    \begin{aligned}
    \text{prox}_{\gamma\mathbf{I}_{np_{1}}, h_{1}}(\gamma^{-1}\text{vec}(\widetilde{\D}_{1}^{(k)}))&=\argmin_{\text{vec}(\Z)\in\mathbb{R}^{np_{1}}} \left[ \frac{1}{2} \|\text{vec}(\Z) - \gamma^{-1}\text{vec}(\widetilde{\D}_{1}^{(k)})\|_{\gamma \mathbf{I}_{np_{1}}}^{2} + h_{1}(\text{vec}(\Z))\right] \\
    &=\argmin_{\text{vec}(\Z)\in\mathbb{R}^{np_{1}}} \left[ \frac{\gamma}{2} \|\text{vec}(\Z) - \gamma^{-1}\text{vec}(\widetilde{\D}_{1}^{(k)})\|_{2}^{2} + h_{1}(\text{vec}(\Z))\right] \\
    &=\argmin_{\Z\in\mathbb{R}^{n\times p_{1}}} \left[ \frac{\gamma}{2} \|\Z - \gamma^{-1}\widetilde{\D}_{1}^{(k)}\|_{F}^{2} + \lambda_{1}\|\Z\|_{*}\right] \\
    &=S(\gamma^{-1}\widetilde{\D}_{1}^{(k)},\gamma^{-1}\lambda_{1}) =\gamma^{-1}S(\widetilde{\D}_{1}^{(k)},\lambda_{1}).     
    \end{aligned}
    \]
    Similarly, the other proximity operators in \eqref{upd:duals} can also be simplified, which leads to the same update of the dual variables in Algorithm \ref{algo:dbfb} of the main paper.
    
    \item For the other part in Algorithm \ref{algo:dbfb} of the main paper, 
    we rewrite
    $x_{t+1} = x_{t} - \sum_{j=1}^{J}\A_{j}^\top\left( y_{t+1}^{j} - y_{t}^{j}\right)$ according to Table \ref{deriv:tab1}. This gives
    \begin{equation}\label{upd:primals}
    \begin{aligned}
    \begin{bmatrix} \text{vec}(\M_{1}^{(t+1)}) \\ \text{vec}(\M_{2}^{(t+1)}) \\ \text{vec}(\M_{3}^{(t+1)}) \end{bmatrix} = & \begin{bmatrix} \text{vec}(\M_{1}^{(t)}) \\ \text{vec}(\M_{2}^{(t)}) \\ \text{vec}(\M_{3}^{(t)}) \end{bmatrix}-\begin{bmatrix} \text{vec}(\D_{1}^{(t+1)}-\D_{1}^{(t)}) \\ \text{vec}(\D_{2}^{(t+1)}-\D_{2}^{(t)}) \\ \text{vec}(\D_{3}^{(t+1)}-\D_{3}^{(t)}) \end{bmatrix} - \begin{bmatrix}
    \text{vec}(\D_{12}^{(t+1)}-\D_{12}^{(t)}) \\
    \mathbf{0}_{np_{3}\times1}
    \end{bmatrix} \\
    -&\begin{bmatrix}
    \text{vec}(\D_{13,1}^{(t+1)}-\D_{13,1}^{(t)}) \\
    \mathbf{0}_{np_{2}\times1} \\
    \text{vec}(\D_{13,2}^{(t+1)}-\D_{13,2}^{(t)})
    \end{bmatrix}- \begin{bmatrix}
    \mathbf{0}_{np_{1}\times1} \\
    \text{vec}(\D_{23}^{(t+1)}-\D_{23}^{(t)})
    \end{bmatrix} \\
    -&\text{vec}(\D^{(t+1)}-\D^{(t)}), 
    \end{aligned}
    \end{equation}
    where $\D_{13,1}^{(t)}\in\mathbb{R}^{n\times p_{1}}$ and $\D_{13,2}^{(t)}\in\mathbb{R}^{n\times p_{3}}$ are submatrices of $\D_{13}^{(t)}=[\D_{13,1}^{(t)}\ \D_{13,2}^{(t)}]$. Thus, we also verify that \eqref{upd:primals} is the same with the update of the primal variables in Algorithm \ref{algo:dbfb} of the main paper.
    
\end{enumerate}

\renewcommand\arraystretch{1.3}%
\setlength{\fboxsep}{0pt}%
\setlength{\fboxrule}{1.5pt}%
\begin{table}[H]
\begin{center}
\fbox{%
  \begin{tabular}{|c|c|}
    \hline
    Abboud et al. (2017) & Algorithm \ref{algo:dbfb} in the main paper \\
    \hline\hline
    $J$ & 7 \\
    \hline
    $\tilde{x}$, $x$, $x_t$ & $\begin{bmatrix} \text{vec}(\X_{1}) \\ \text{vec}(\X_{2})  \\ \text{vec}(\X_{3}) \end{bmatrix}$, $\begin{bmatrix} \text{vec}(\M_{1}) \\ \text{vec}(\M_{2}) \\ \text{vec}(\M_{3}) \end{bmatrix}$, $\begin{bmatrix} \text{vec}(\M_{1}^{(t)}) \\ \text{vec}(\M_{2}^{(t)}) \\ \text{vec}(\M_{3}^{(t)}) \end{bmatrix}$ \\
    \hline    
    $\mathbf{A}_{1}$, $h_{1}(\A_{1}x)$ & $\left[\mathbf{I}_{ np_{1}\times np_{1}} \quad \mathbf{0}_{np_{1} \times np_{2}}\quad \mathbf{0}_{np_{1} \times np_{3}} \right]$, $\lambda_{1}\|\M_{1}\|_{*}$ \\    
    \hline
    $\mathbf{A}_{2}$, $h_{2}(\A_{2}x)$ & $\left[\mathbf{0}_{ np_{2}\times np_{1}} \quad \mathbf{I}_{np_{2} \times np_{2}}\quad \mathbf{0}_{np_{2} \times np_{3}} \right]$, $\lambda_{2}\|\M_{2}\|_{*}$ \\    
    \hline
    $\mathbf{A}_{3}$, $h_{3}(\A_{3}x)$ & $\left[\mathbf{0}_{ np_{3}\times np_{1}} \quad \mathbf{0}_{np_{3} \times np_{2}}\quad \mathbf{I}_{np_{3} \times np_{3}} \right]$, $\lambda_{3}\|\M_{3}\|_{*}$ \\    
    \hline 
    $\mathbf{A}_{4}$, $h_{4}(\A_{4}x)$ & $\begin{bmatrix} \mathbf{I}_{ np_{1}\times np_{1}} \quad \mathbf{0}_{np_{1} \times np_{2}} \quad \mathbf{0}_{np_{1} \times np_{3}} \\
    \mathbf{0}_{ np_{2}\times np_{1}} \quad \mathbf{I}_{np_{2} \times np_{2}} \quad \mathbf{0}_{np_{2} \times np_{3}}
    \end{bmatrix}$, $\lambda_{12}\|\M_{12}\|_{*}$ \\     
    \hline
    $\mathbf{A}_{5}$, $h_{5}(\A_{5}x)$ & $\begin{bmatrix} \mathbf{I}_{ np_{1}\times np_{1}} \quad \mathbf{0}_{np_{1} \times np_{2}} \quad \mathbf{0}_{np_{1} \times np_{3}} \\
    \mathbf{0}_{ np_{3}\times np_{1}} \quad \mathbf{0}_{np_{3} \times np_{2}} \quad \mathbf{I}_{np_{3} \times np_{3}}
    \end{bmatrix}$, $\lambda_{13}\|[\M_{13}\|_{*}$ \\     
    \hline
    $\mathbf{A}_{6}$, $h_{6}(\A_{6}x)$ &   $\begin{bmatrix} \mathbf{0}_{ np_{2}\times np_{1}} \quad \mathbf{I}_{np_{2} \times np_{2}} \quad \mathbf{0}_{np_{2} \times np_{3}} \\
    \mathbf{0}_{ np_{3}\times np_{1}} \quad \mathbf{0}_{np_{3} \times np_{2}} \quad \mathbf{I}_{np_{3} \times np_{3}}
    \end{bmatrix}$, $\lambda_{23}\|\M_{23}\|_{*}$ \\      
    \hline 
    $\mathbf{A}_{7}$, $h_{7}(\A_{7}x)$ &  $\I_{n(p_{1}+p_{2}+p_{3})\times n(p_{1}+p_{2}+p_{3})}$, $\lambda\|\M\|_{*}$ \\    
    \hline
    $\mathbf{B}_{1}$, $\mathbf{B}_{2}$, $\mathbf{B}_{3}$, $\mathbf{B}_{4}$, $\mathbf{B}_{5}$, $\mathbf{B}_{6}$, $\mathbf{B}_{7}$ & $\mathbf{I}_{np_{1} }$, $\mathbf{I}_{np_{2}}$, $\mathbf{I}_{np_{3}}$, $\mathbf{I}_{n(p_{1}+p_{2})}$, $\mathbf{I}_{n(p_{1}+p_{3})}$, $\mathbf{I}_{n(p_{2}+p_{3})}$, $\mathbf{I}_{n(p_{1}+p_{2}+p_{3})}$ \\
    \hline
    $y_{t+1}^{1}$, $\tilde{y}_{t}^{1}$ & $\text{vec}(\D_{1}^{(t+1)}),\text{vec}(\D_{1}^{(t)}+\gamma\M_{1}^{(t)})\in\mathbb{R}^{np_{1}\times1}$ \\
    \hline
    $y_{t+1}^{2}$, $\tilde{y}_{t}^{2}$ & $\text{vec}(\D_{2}^{(t+1)}),\text{vec}(\D_{2}^{(t)}+\gamma\M_{2}^{(t)})\in\mathbb{R}^{np_{2}\times1}$ \\
    \hline
    $y_{t+1}^{3}$, $\tilde{y}_{t}^{3}$ & $\text{vec}(\D_{3}^{(t+1)}),\text{vec}(\D_{3}^{(t)}+\gamma\M_{3}^{(t)})\in\mathbb{R}^{np_{3}\times1}$ \\
    \hline
    $y_{t+1}^{4}$, $\tilde{y}_{t}^{4}$ & $\text{vec}(\D_{12}^{(t+1)}),\text{vec}(\D_{12}^{(t)}+\gamma\M_{12}^{(t)})\in\mathbb{R}^{n(p_{1}+p_{2})\times1}$ \\
    \hline
    $y_{t+1}^{5}$, $\tilde{y}_{t}^{5}$ & $\text{vec}(\D_{13}^{(t+1)}),\text{vec}(\D_{13}^{(t)}+\gamma\M_{13}^{(t)})\in\mathbb{R}^{n(p_{1}+p_{3})\times1}$ \\
    \hline
    $y_{t+1}^{6}$, $\tilde{y}_{t}^{6}$ & $\text{vec}(\D_{23}^{(t+1)}),\text{vec}(\D_{23}^{(t)}+\gamma\M_{23}^{(t)})\in\mathbb{R}^{n(p_{2}+p_{3})\times1}$ \\
    \hline
    $y_{t+1}^{7}$, $\tilde{y}_{t}^{7}$ & $\text{vec}(\D^{(t+1)}),\text{vec}(\D^{(t)}+\gamma\M^{(t)})\in\mathbb{R}^{n(p_{1}+p_{2}+p_{3})\times1}$ \\
    \hline
  \end{tabular}%
 }
 \end{center}
\caption{Each column shows the corresponding notation in \cite{abboud2017dual} and our work. The notation in the same row corresponds to each other. $\text{vec}(\cdot)$ denotes vectorization of matrix (i.e., stacking the columns of the input matrix into a single vector). The subscripts of $\I$ and $\mathbf{0}$ indicate the dimensions of the identity and zero matrices, respectively.}
\label{deriv:tab1}
\end{table}

\subsection{Empirical result of refitting procedure} 

\begin{figure}[H]
\centering
\includegraphics[scale = 0.8]{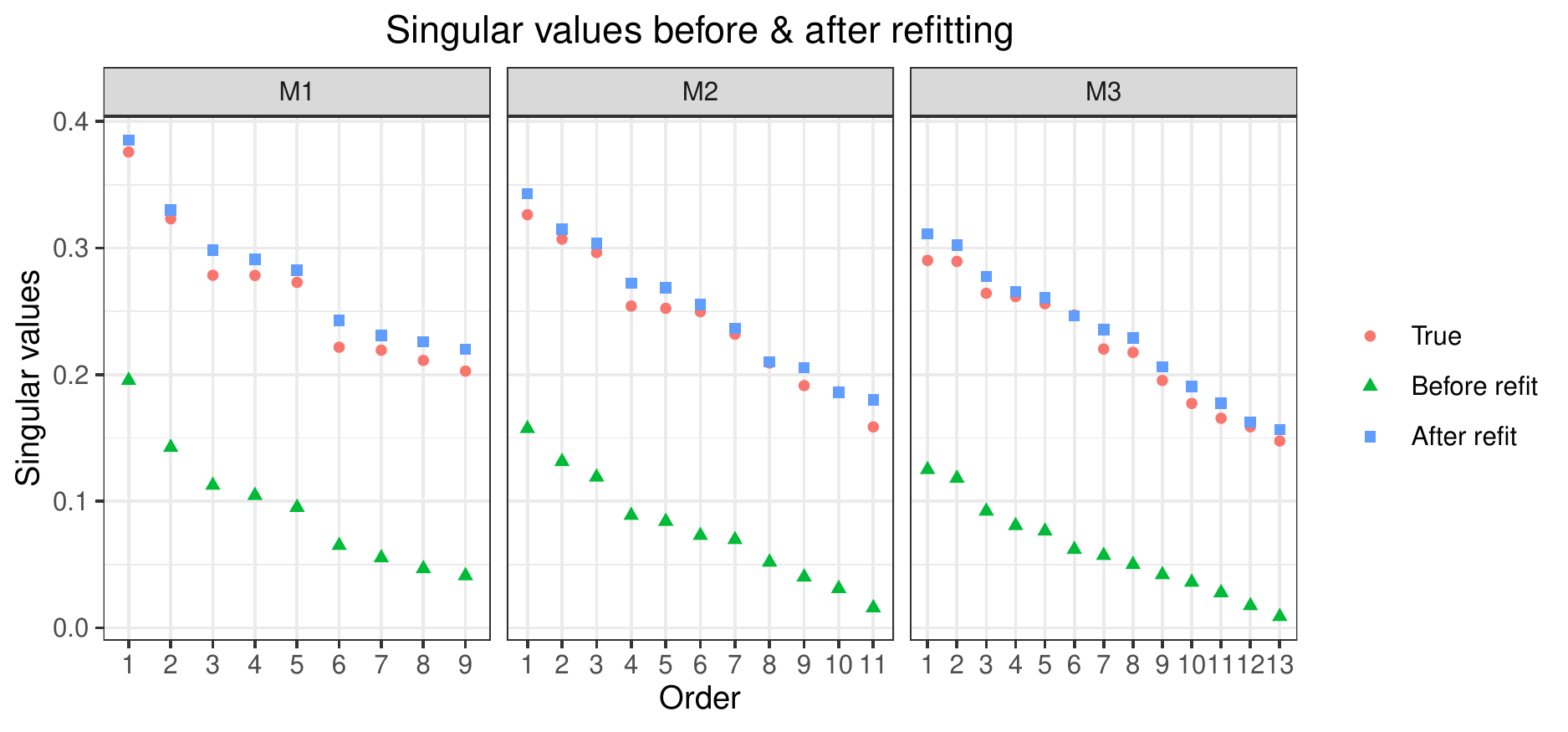}
\caption{Scatter plots of singular values of $\widehat{\M}_{d}$ (green triangles), $\widehat{\M}_{d}^{\text{refit}}$ (blue squares) and the true signal $\M_{d}$ (red circles) for $d=1,2,3$. The figure is based on one replication from our additional simulation studies in \ref{supp:simres} with the tuning parameter chosen by BCV in Section \ref{subsec:bcv}}\label{fig:refit}
\end{figure}

Figure \ref{fig:refit} illustrates the performance of our refitting method to adjust bias of singular values.

\subsection{Setting up tuning grid}

Recall that our modified hierarchical nuclear norm penalties are:
\begin{equation}\label{pen:mod}
\begin{aligned}
&\text{$D=2$ views:}\quad \tau\sum_{d=1}^{3} \omega_{d}^{\text{SURE}}\|\M_{d}\|_{*} +\lambda\|\M\|_{*}  \\
&\text{$D=3$ views:}\quad \tau\sum_{d=1}^{3} \omega_{d}^{\text{SURE}}\|\M_{d}\|_{*} + \kappa\sum_{k,l=1,k<l}^{3}\omega_{kl}^{\text{SURE}}\|\M_{kl}\|_{*} +\lambda\|\M\|_{*} 
\end{aligned}
\end{equation}    
where $\tau,\kappa>0$ control the overall shrinkage amount for the individual and pairwise matrices, respectively, and the relative weights are chosen as
\begin{equation}\label{sure:wts}
\omega_{d}^{\text{SURE}} = \frac{\lambda_{d}^{\text{SURE}}}{\lambda_{1}^{\text{SURE}} + \lambda_{2}^{\text{SURE}} + \lambda_{3}^{\text{SURE}}}, \quad \omega_{kl}^{\text{SURE}} = \frac{\lambda_{kl}^{\text{SURE}}}{\lambda_{12}^{\text{SURE}} + \lambda_{13}^{\text{SURE}} + \lambda_{23}^{\text{SURE}}}.
\end{equation}
In \eqref{sure:wts}, $\lambda_{d}^{\text{SURE}}$ denotes the optimal cutoff of $S(\X_{d},\lambda_{d})$ chosen by the the Stein's Unbiased Risk Estimates (SURE) \citep{candes2013unbiased} such that 
\begin{equation}\label{sure:min}
\lambda_{d}^{\text{SURE}}=\argmin_{\lambda_{d}}\text{SURE}(S(\X_{d},\lambda_{d})), \end{equation}
where $\text{SURE}(S(\X_{d},\lambda_{d}))$ is unbiased risk estimates with $\text{div}\left(S(\X_{d},\lambda_{d})\right)$ as the degrees of freedom estimates such that 
\begin{equation}\label{sure:form}
\begin{aligned}
\text{SURE}(S(\X_{d},\lambda_{d}))=&-np_{d}s_{d}^{2}+\sum_{i=1}^{\min(n,p_{d})}\min(\lambda_{d}^{2},\sigma_{i}(\X_{d})^{2}) + 2s_{d}^{2}\text{div}\left(S(\X_{d},\lambda_{d})\right),\\
\text{div}\left(S(\X_{d},\lambda_{d})\right)=&|n-p_{d}|\sum_{i=1}^{\min(n,p_{d})}\left(1-\frac{\lambda_{d}}{\sigma_{i}(\X_{d})}\right)_{+}+\sum_{i=1}^{\min(n,p_{d})}\mathbf{1}\left\{\sigma_{i}(\X_{d})>\lambda_{d} \right\} \\
&+2\sum_{i\neq j, i,j=1}^{\min(n,p_{d})} \frac{\sigma_{i}(\X_{d})(\sigma_{i}(\X_{d})-\lambda_{d})_{+}}{\sigma_{i}(\X_{d})^{2}-\sigma_{j}(\X_{d})^{2}},
\end{aligned}
\end{equation}
and the noise level $s_{d}$ comes from the underlying model $\X_{d}=\M_{d}+\E_{d}$ with each element of noise matrix $\E_{d}$ from $N(0,s_{d}^{2})$. Similarly, $\lambda_{kl}^{\text{SURE}}$ is the optimal cutoff of $S(\X_{kl},\lambda_{kl})$ chosen by the SURE criteria. In practice, calculating the SURE criteria \eqref{sure:form} requires the noise level as an input for which we use the median absolute deviation (MAD) estimator \citep{gavish2017optimal} by using the R package \texttt{denoiseR} \citep{josse2016denoiser}.

To set up the tuning grid, we first calculate the upper bounds $\tau_{\text{max}}$, $\kappa_{\text{max}}$ and $\lambda_{\text{max}}$ for each tuning parameter such that
\[
S(\X_{d},\tau_{\text{max}})=\mathbf{0},\quad S(\X_{kl},\kappa_{\text{max}})=\mathbf{0},\quad S(\X,\lambda_{\text{max}})=\mathbf{0},
\]
that is, any singular values are shrunk to be zero when the cutoffs $\tau$, $\kappa$ and $\lambda$ are greater than $\tau_{\text{max}}$, $\kappa_{\text{max}}$ and $\lambda_{\text{max}}$. Those can be found as follows:

\begin{itemize}
    
    \item $\tau_{\text{max}}$: Set $\kappa=0$ and $\lambda=0$. For any $\tau\frac{\lambda_{d}^{\text{SURE}}}{\lambda_{1}^{\text{SURE}} + \lambda_{2}^{\text{SURE}} + \lambda_{3}^{\text{SURE}}}\geq\sigma_{\text{max}}(\X_{d})$, we have $S(\X_{d},\tau)=\mathbf{0}$ for $d=1,2,3$, which implies
    \[
    \tau_{\text{max}} = \text{max}\left\{\frac{\sigma_{\text{max}}(\X_{1})}{\lambda_{1}^{\text{SURE}}}, \frac{\sigma_{\text{max}}(\X_{2})}{\lambda_{2}^{\text{SURE}}}, \frac{\sigma_{\text{max}}(\X_{3})}{\lambda_{3}^{\text{SURE}}} \right\}\sum_{d=1}^{3}\lambda_{d}^{\text{SURE}}.
    \]
    
    \item $\kappa_{\text{max}}$: Set $\tau=0$ and $\lambda=0$. Similarly, for any $\kappa\frac{\lambda_{kl}^{\text{SURE}}}{\lambda_{12}^{\text{SURE}} + \lambda_{13}^{\text{SURE}} + \lambda_{23}^{\text{SURE}}}\geq\sigma_{\text{max}}(\X_{kl})$, we have $S(\X_{kl},\kappa)=\mathbf{0}$ for $(k,l)=(1,2),(1,3),(2,3)$, which implies
    \[
    \kappa_{\text{max}} = \text{max}\left\{\frac{\sigma_{\text{max}}(\X_{12})}{\lambda_{12}^{\text{SURE}}}, \frac{\sigma_{\text{max}}(\X_{13})}{\lambda_{13}^{\text{SURE}}}, \frac{\sigma_{\text{max}}(\X_{23})}{\lambda_{23}^{\text{SURE}}} \right\}\sum_{k,l=1,k<l}^{3} \lambda_{kl}^{\text{SURE}}.
    \]

    \item $\lambda_{\text{max}}$: Set $\tau=0$ and $\kappa=0$.It is clear that $\lambda_{\text{max}}=\sigma_{\text{max}}(\X)$ satisfies $S(\X,\lambda_{\text{max}})=\mathbf{0}$.    
\end{itemize}

Given $\tau_{\text{max}}$, $\kappa_{\text{max}}$ and $\lambda_{\text{max}}$, we set up the tuning grid of $\left\{(\tau_{l},\kappa_{l},\lambda_{l})\right\}_{l=1}^{L}$ as follows:

\begin{enumerate}

    \item[]Step 1: Create three equally-spaced grids on a logarithmic scale from -5 to $\log\tau_{\text{max}}$, $\log\kappa_{\text{max}}$ and $\log\lambda_{\text{max}}$ with length 10. $\tau=\kappa=\lambda=0$ is also included in each sequence. In \texttt{R}, we can use the following code:   
    \[
    \begin{aligned}
    &\text{tau\_seq: \texttt{c(0, exp(seq(-5, log\_tau\_max, len = 10)))}}, \\
    &\text{kappa\_seq: \texttt{c(0, exp(seq(-5, log\_kappa\_max, len = 10)))}}, \\
    &\text{lambda\_seq: \texttt{c(0, exp(seq(-5, log\_lambda\_max, len = 10)))}}. \\    
    \end{aligned}
    \]
    
    \item[]Step 2: Consider all combinations of each grid points of $\tau$, $\kappa$ and $\lambda$, that is, \[
    G=\left\{(\tau,\kappa,\lambda):\tau\in S_{\tau},\kappa\in S_{\kappa},\lambda\in S_{\lambda} \right\}
    \]
    where $S_{\tau}$, $S_{\kappa}$ and $S_{\lambda}$ denote the set of the above three sequences of $\tau$, $\kappa$ and $\lambda$.

    \item[]Step 3: Consider the hyperplane $H(\tau,\kappa,\lambda): a\tau+b\kappa+c\lambda=d$ passing through $(\tau_{\text{max}},0,0)$, $(0,\kappa_{\text{max}},0)$ and $(0,0,\lambda_{\text{max}})$. Given the first two coordinates $(\tau,\kappa)$, we let $h(\tau,\kappa)$ be the value of the third coordinate such that $(\tau,\kappa,h(\tau,\kappa))$ lies on that hyperplane. Then, our tuning grid is chosen as
    \[
    \left\{(\tau_{l},\kappa_{l},\lambda_{l})\right\}_{l=1}^{L}=\left\{(\tau,\kappa,\lambda): \lambda\leq h(\tau,\kappa) \text{ for each } (\tau,\kappa,\lambda)\in G \right\}.
    \]
    That is, we only keep the elements in the set $G$ when the corresponding value of the 3rd coordinate is less than or equal to $h(\tau,\kappa)$. Thus, the resulting $\left\{(\tau_{l},\kappa_{l},\lambda_{l})\right\}_{l=1}^{L}$ would be a tetrahedron. This is motivated to avoid the case where all singular values are truncated to be zero.
\end{enumerate}

\subsection{One standard error rule for choosing tuning parameters}

Given the grid of tuning parameters $\left\{(\tau_{l}, \kappa_{l}, \lambda_{l}) \right\}_{l=1}^{L}$, let $\text{Avg\_Err}_{l} = 4^{-1}\sum_{j,k=1}^{2}\text{BCVErr}_{j,k,l}$ be the average BCV error for combination $l$, and let $\text{Avg\_Err}_{\text{min}}=\min_{1\leq l \leq L} \text{Avg\_Err}_{l}$ be the minimal error. Let $\widehat{\text{SE}}(\text{Avg\_Err}_{\text{min}})$ be the standard error estimate of $\text{Avg\_Err}_{\text{min}}$. To choose the most parsimonious model in terms of total rank of the estimated concatenated $\widehat{\M}^{\text{refit}}$, we propose to pick the optimal $(\tau^{*}, \kappa^{*}, \lambda^{*})$ based on 1 standard error rule \citep{hastie2009elements}. 
Specifically, let $\text{rank}\{\widehat{\M}^{\text{refit}}(\tau_{l},\kappa_{l},\lambda_{l})\}$ denote the total rank of the concatenated $\widehat{\M}^{\text{refit}}$  obtained using full data $\X$ with parameters $(\tau_{l},\kappa_{l},\lambda_{l})$. We determine optimal tuning parameters $(\tau^{*}, \kappa^{*}, \lambda^{*})$ as
\begin{equation}\label{1se}
\begin{aligned}
&(\tau^{*}, \kappa^{*}, \lambda^{*})=\argmin_{\tau_{l},\kappa_{l},\lambda_{l}}\ \text{rank}\{\widehat{\M}^{\text{refit}}(\tau_{l},\kappa_{l},\lambda_{l})\} \\
&\text{subject to}\quad \text{Avg\_Err}_{l}\leq \text{Avg\_Err}_{\text{min}} + \widehat{\text{SE}}(\text{Avg\_Err}_{\text{min}}),
\end{aligned}
\end{equation}
where $\widehat{\text{SE}}(\text{Avg\_Err}_{\text{min}})$ is the standard error estimates of $\text{Avg\_Err}_{\text{min}}$ such that 
\[
\widehat{\text{SE}}(\text{Avg\_Err}_{\text{min}})=\sqrt{\frac{1}{4}\sum_{j,k=1}^{2}\frac{\{\text{BCVErr}_{j,k,\text{min}}- \text{Avg\_Err}_{\text{min}}\}^{2}}{4-1}},
\]
where $\text{BCVErr}_{j,k,\text{min}}$ denotes the BCV error of the held-out $\X_{d}^{j,k}$ at the value of the tuning parameter that minimizes the average BCV error.
In \eqref{1se}, we consider all $(\tau_{l}, \kappa_{l}, \lambda_{l})$ whose corresponding average BCV error is at most one standard error larger than the minimal error, and select among them $(\tau^{*}, \kappa^{*}, \lambda^{*})$ as the one that minimizes total rank on the full dataset.

\section{Supplementary material for Section \ref{sec:sim}}\label{supp:sim}

\subsection{Implementation details}

We compare the performance of our approach with the following methods:

\begin{enumerate}
    
    \item JIVE \citep{lock2013joint} using the R package \texttt{r.jive} \citep{o2016r}
    with (i) the default permutation method for rank selection and (ii) the true ranks

    \item AJIVE \citep{feng2018angle} using the MATLAB code available at \url{https://github.com/MeileiJiang/AJIVE_Project} with the initial ranks chosen by the R package \texttt{igraph} using the profile likelihood approach \citep{zhu2006automatic}
    
    \item SLIDE \citep{gaynanova2019structural} using the R package \texttt{SLIDE} (\url{https://github.com/irinagain/SLIDE}) with their adapted $2\times2$ BCV scheme for rank selection
    
    \item BIDIFAC \citep{park2020integrative} and BIDIFAC+ \citep{lock2022bidimensional}. We use R code available at \url{https://github.com/lockEF/bidifac} to implement BIDIFAC+, and modify this code for multi-view case to implement BIDIFAC. Since neither approach makes a bias adjustment to account for penalization, 
    we also apply our refitting step (Algorithm \ref{algo:refit}) with both methods for fair comparison. The corresponding results are labeled as BIDIFAC\_refit and BIDIFAC+\_refit.
    
\end{enumerate}

For HNN, JIVE, AJIVE and SLIDE, each $\X_{d}$ is first column-centered and scaled so that $\|\X_{d}\|_{F}=1$ in agreement with the standard preprocessing of data matrices from multiple views \citep{smilde2003framework,lock2013joint}. For BIDIFAC and BIDIFAC+, the scaling is done as recommended in \citet{park2020integrative,lock2022bidimensional} by using MAD estimator \citep{gavish2017optimal} of noise level.

\subsection{Data generation}

\begin{enumerate}
    \item[(a)] \textit{Two views} ($D=2$).
    We set $n=150$ and $p_{1}=p_{2}=50$. The signal matrices $\M_1$ and $\M_2$ are generated in the following way with the joint and individual ranks as $s_{0}=2$ and $s_{1}=s_{2}=4$: 
    \[
    \begin{aligned}
    \M_{1}&=\U_{0}\D_{0}\V_{1,0}^\top + \U_{1}\D_{1}\V_{1,1}^\top \in\mathbb{R}^{n \times p_{1}}, \\
    \M_{2}&= \U_{0}\D_{0}\V_{2,0}^\top + \U_{2}\D_{2}\V_{2,2}^\top \in\mathbb{R}^{n \times p_{2}}, \\
    \end{aligned}
    \]
    where the elements of the scores $\U_{0}\in\mathbb{R}^{n \times s_{0}}$, $\U_{1}\in\mathbb{R}^{n \times s_{1}}$, $\U_{2}\in\mathbb{R}^{n \times s_{2}}$ and the loadings $\V_{1,0} \in\mathbb{R}^{p_{1} \times s_{0}},\V_{2,0} \in\mathbb{R}^{ p_{2} \times s_{0}},\V_{1,1} \in\mathbb{R}^{p_{1} \times s_{1}},\V_{2,2} \in\mathbb{R}^{p_{2} \times s_{2}}$ are drawn independently from $\text{Unif}(0,1)$ with subsequent orthonormalization of each $\U_{0},\U_{1},\U_{2}$ and $\V=\begin{bmatrix}
    \V_{1,0} & \V_{1,1} & \mathbf{0} \\
    \V_{2,0} & \mathbf{0} & \V_{2,2} \\    
    \end{bmatrix}.$ For the individual scores, we consider the following two cases: 
    \begin{enumerate}
       \item[(i)] Orthogonal: $\mathcal{C}(\U_{1})\perp\mathcal{C}(\U_{2})$ 
       \item[(ii)] Non-orthogonal: $\mathcal{C}(\U_{1})\not\perp\mathcal{C}(\U_{2})$. Specifically, the four principal angles between $\mathcal{C}(\U_{1})$ and $\mathcal{C}(\U_{2})$ are set to be approximately
        $30^{\circ},40^{\circ},50^{\circ},60^{\circ}$        
    \end{enumerate}
    Given $\U_{1}$ and $\U_{2}$, the joint structure $\mathcal{C}(\U_{0})$ is set to be orthogonal to both individual spaces. The elements of the diagonal matrices $\D_{0}\in\mathbb{R}^{ s_{0} \times s_{0}},\D_{1}\in\mathbb{R}^{ s_{1} \times s_{1}},\D_{2}\in\mathbb{R}^{ s_{2} \times s_{2}}$ are drawn independently from $\text{Unif}(1.0, 1.5)$. The noise matrix is generated such that $\text{SNR}=1$.
    
    \item[(b)] \textit{Three views} ($D=3$). We set $n=100$ and $p_{1}=p_{2}=p_{3}=100$.
    The corresponding signal matrices $\M_1$, $\M_2$ and $\M_3$ are generated with the ranks $s_{0}=s_{23}=s_{1}=s_{2}=s_{3}=2$ and $s_{12}=s_{13}=4$ in the following way:
    \begin{equation}\label{dgen:d3}
    \begin{aligned}
    \M_{1}&=\U_{0}\D_{0}\V_{1,0}^\top + \U_{12}\D_{12}\V_{1,12}^\top + \U_{13}\D_{13}\V_{1,13}^\top + \U_{1}\D_{1}\V_{1,1}^\top \in\mathbb{R}^{n \times p_{1}}, \\
    \M_{2}&= \U_{0}\D_{0}\V_{2,0}^\top + \U_{12}\D_{12}\V_{2,12}^\top + \U_{23}\D_{23}\V_{2,23}^\top + \U_{2}\D_{2}\V_{2,2}^\top\in\mathbb{R}^{n \times p_{2}}, \\
    \M_{3}&= \U_{0}\D_{0}\V_{3,0}^\top + \U_{13}\D_{13}\V_{3,13}^\top + \U_{23}\D_{23}\V_{3,23}^\top + \U_{3}\D_{3}\V_{3,3}^\top\in\mathbb{R}^{n \times p_{3}}, \\
    \end{aligned}        
    \end{equation}
    where the elements of the scores ($\U_{0}\in\mathbb{R}^{n \times s_{0}}$, $\U_{1}\in\mathbb{R}^{n \times s_{1}}$, $\U_{2}\in\mathbb{R}^{n \times s_{2}}$, $\U_{3}\in\mathbb{R}^{n \times s_{3}}$, $\U_{12}\in\mathbb{R}^{n \times s_{12}}$, $\U_{13}\in\mathbb{R}^{n \times s_{13}}$, $\U_{23}\in\mathbb{R}^{n \times s_{23}}$), loadings ($\V_{d,0} \in\mathbb{R}^{p_{d} \times s_{0}},\V_{d,12} \in\mathbb{R}^{ p_{d} \times s_{12}},\V_{d,13} \in\mathbb{R}^{ p_{d} \times s_{13}},\V_{d,23} \in\mathbb{R}^{p_{d} \times s_{23}},\V_{d,d} \in\mathbb{R}^{p_{d} \times s_{d}}$ for $d=1,2,3$) and diagonal matrices ($\D_{0}\in\mathbb{R}^{ s_{0} \times s_{0}},\D_{1}\in\mathbb{R}^{ s_{1} \times s_{1}},\D_{2}\in\mathbb{R}^{ s_{2} \times s_{2}},\D_{3}\in\mathbb{R}^{ s_{3} \times s_{3}},\D_{12}\in\mathbb{R}^{ s_{12} \times s_{12}},\D_{13}\in\mathbb{R}^{ s_{13} \times s_{13}},\D_{23}\in\mathbb{R}^{ s_{23} \times s_{23}}$) are generated as in (a). For both partially-shared scores $\U_{12}$ and $\U_{13}$, we consider the following two cases: 
    \begin{enumerate}
        \item[(i)] Orthogonal: $\mathcal{C}(\U_{12})\perp\mathcal{C}(\U_{13})$
        \item[(ii)] Non-orthogonal: $\mathcal{C}(\U_{12})\not\perp\mathcal{C}(\U_{13})$. Specifically, the four principal angles between $\mathcal{C}(\U_{12})$ and $\mathcal{C}(\U_{13})$ are set to be approximately $30^{\circ},40^{\circ},50^{\circ},60^{\circ}$       
    \end{enumerate}
    Given $\mathcal{C}(\U_{12})$ and $\mathcal{C}(\U_{13})$, the other signal structures induced by each score matrices are set to be orthogonal to each other. The noise matrix is generated such that $\text{SNR}=2$.    
\end{enumerate}

\subsection{Results with different BCV splits} 

\begin{figure}[H]
    \centering
    \includegraphics[scale = 0.75]{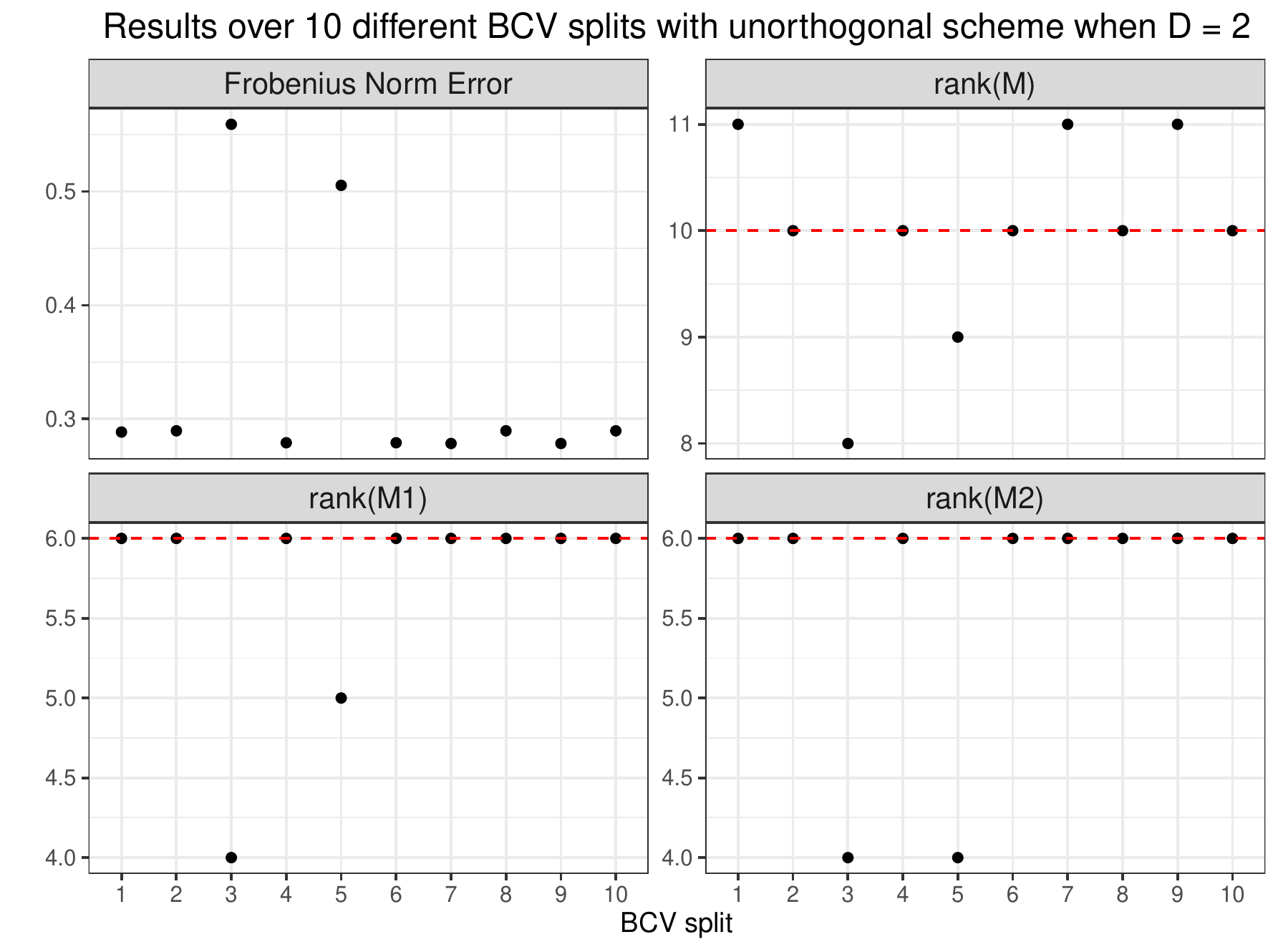} 
    \caption{Each panel shows scaled Frobenius norm error (Top Left), rank estimates of $\M$ (Top right), $\M_{1}$ (Bottom left) and $\M_{2}$ (Bottom right) from HNN by the one-SE rule over 10 different $2\times2$ BCV splits on the same simulated data with the unorthogonal scheme when $D=2$. The true ranks are indicated by the red dotted line}\label{fig:split}
\end{figure}

Figure \ref{fig:split} shows the results of our HNN estimator based on the one SE rule with 10 different $2\times2$ BCV splits of the same simulated data, for the unorthogonal settings when $D=2$. While there are two splits (the 3rd and 5th splits) showing large Frobenius norm error and smaller rank estimates than the true rank, it is clear that the other BCV splits show smaller Frobenius norm errors with more precise rank estimates. 

\section{Additional simulation results}\label{supp:simres}

We present additional simulation results for $D=3$ in this section. We generate signal matrices by \eqref{dgen:d3} with $n=100$, $p_{1}=p_{2}=p_{3}=100$ and the two rank schemes: (a) $s_{0}=s_{1}=s_{2}=s_{3}=s_{12}=s_{13}=s_{23}=2$; (b) $s_{0}=1, s_{1}=s_{2}=s_{3}=2, s_{12}=1, s_{13}=3, s_{23}=5$. To simulate the score matrices, we consider the following two cases: 
\begin{enumerate}
    \item[] (i) Orthogonal: $\U^\top\U=\I$ where $\U=[\U_{0}\ \U_{12}\ \U_{13}\ \U_{23}\ \U_{1}\ \U_{2}\ \U_{3} ]$
    \item[] (ii) Non-orthogonal: $\mathcal{C}(\U_{12})\not\perp\mathcal{C}(\U_{13})$, $\mathcal{C}(\U_{12})\not\perp\mathcal{C}(\U_{23})$, $\mathcal{C}(\U_{13})\not\perp\mathcal{C}(\U_{23})$ and $\mathcal{C}(\U_{j})\not\perp\mathcal{C}(\U_{kl})$ for distinct $j,k,l$. Any other pairs are orthogonal.
\end{enumerate}
The elements of the loading and singular values are generated as in \eqref{dgen:d3}. The noise matrix is generated such that $\text{SNR}=2$.   

\begin{figure}[H]
    \centering
    \includegraphics[scale = 0.68]{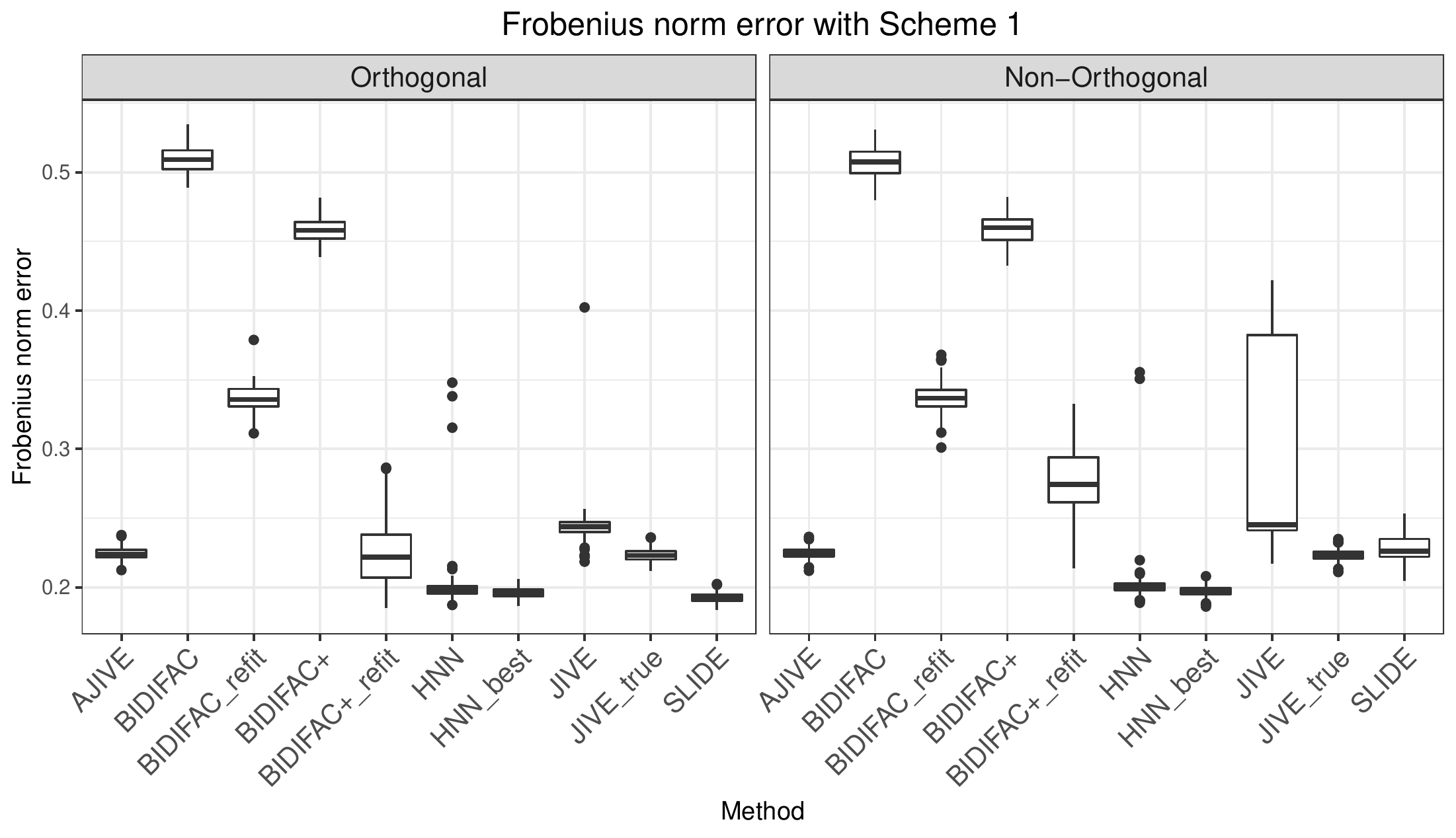}         
    \includegraphics[scale = 0.68]{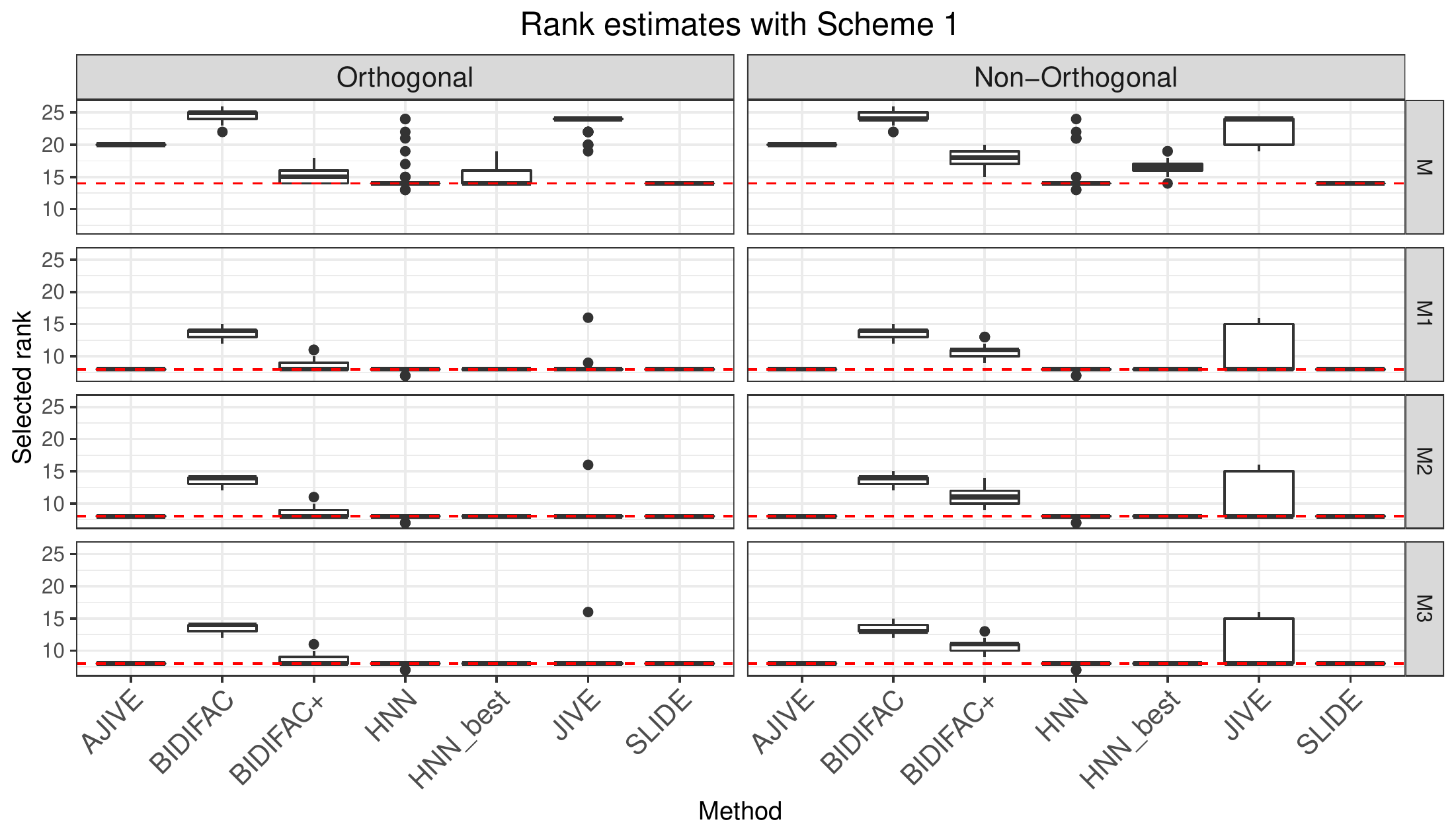}         
    \caption{Boxplots of scaled squared Frobenius norm errors \eqref{sfb} and rank estimates of the concatenated signals and each $\M_{d}$ over 100 independent replication for Orthogonal scheme (left column) and Non-orthogonal scheme (right column) when $D = 3$ with the ranks $s_{0}=s_{1}=s_{2}=s_{3}=s_{12}=s_{13}=s_{23}=2$. The true ranks are indicated by the red dotted line.}\label{sim:fig1_asim_d3}    
\end{figure}

\begin{figure}[H]
    \centering
    \includegraphics[scale = 0.68]{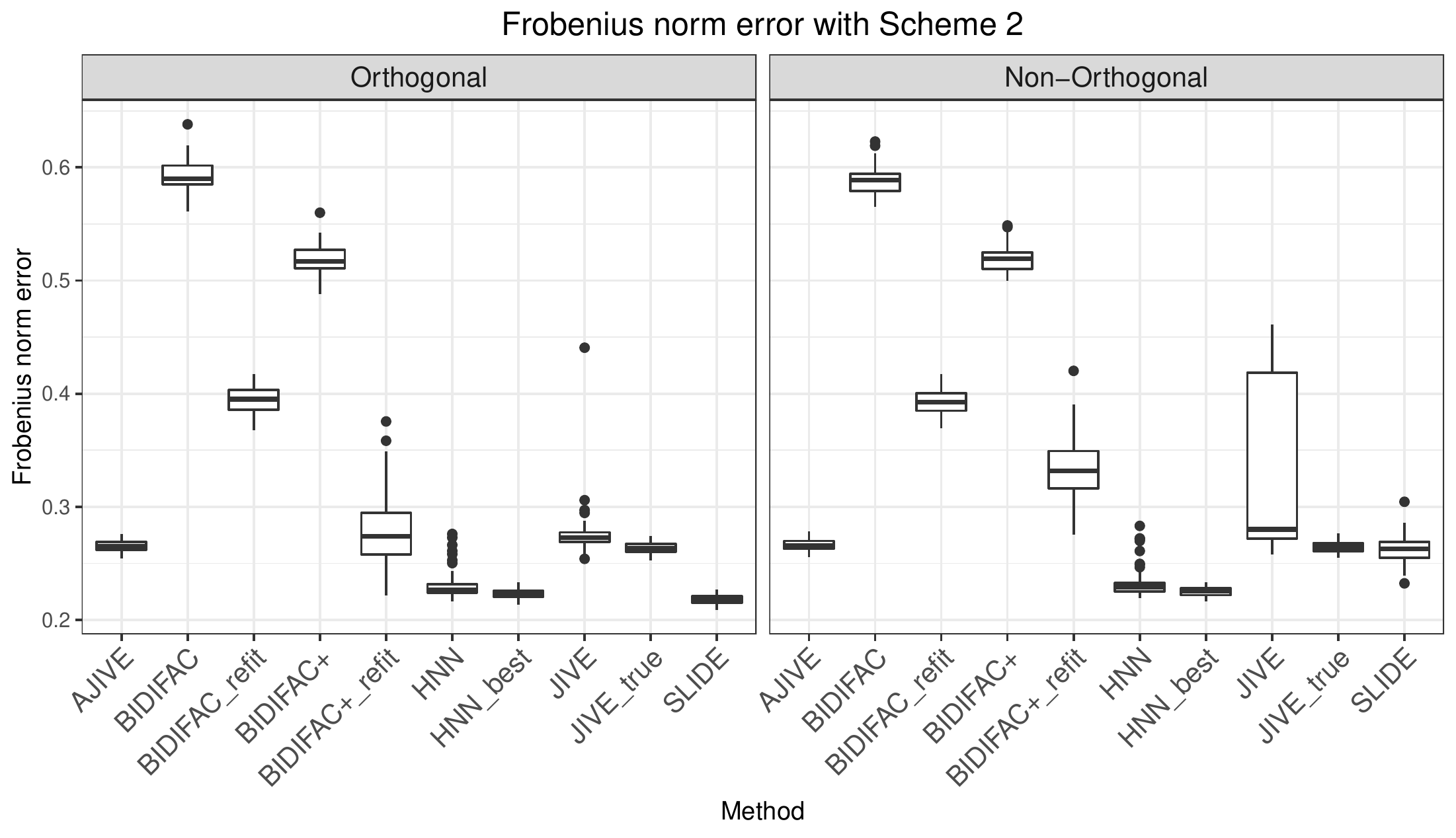}         
    \includegraphics[scale = 0.68]{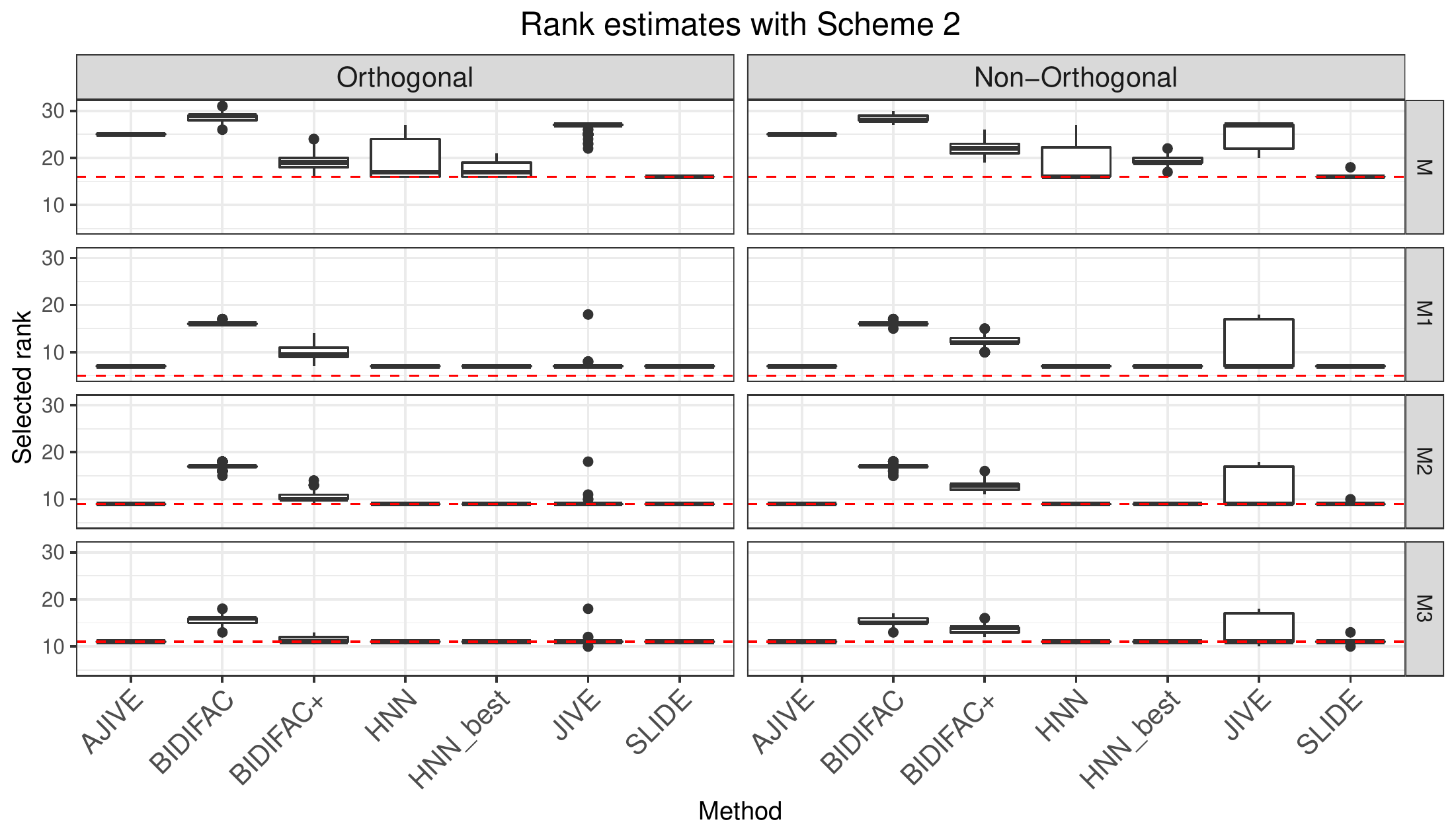}         
    \caption{Boxplots of scaled squared Frobenius norm errors \eqref{sfb} and rank estimates of the concatenated signals and each $\M_{d}$ over 100 independent replication for Orthogonal scheme (left column) and Non-orthogonal scheme (right column) when $D = 3$ with the ranks $s_{0}=1, s_{1}=s_{2}=s_{3}=2, s_{12}=1, s_{13}=3, s_{23}=5$. The true ranks are indicated by the red dotted line.}\label{sim:fig2_asim_d3}    
\end{figure}

\section{Additional analysis of GTEx data}\label{supp:real}

\begin{figure}[H]
    \centering
    \includegraphics[scale = 0.9]{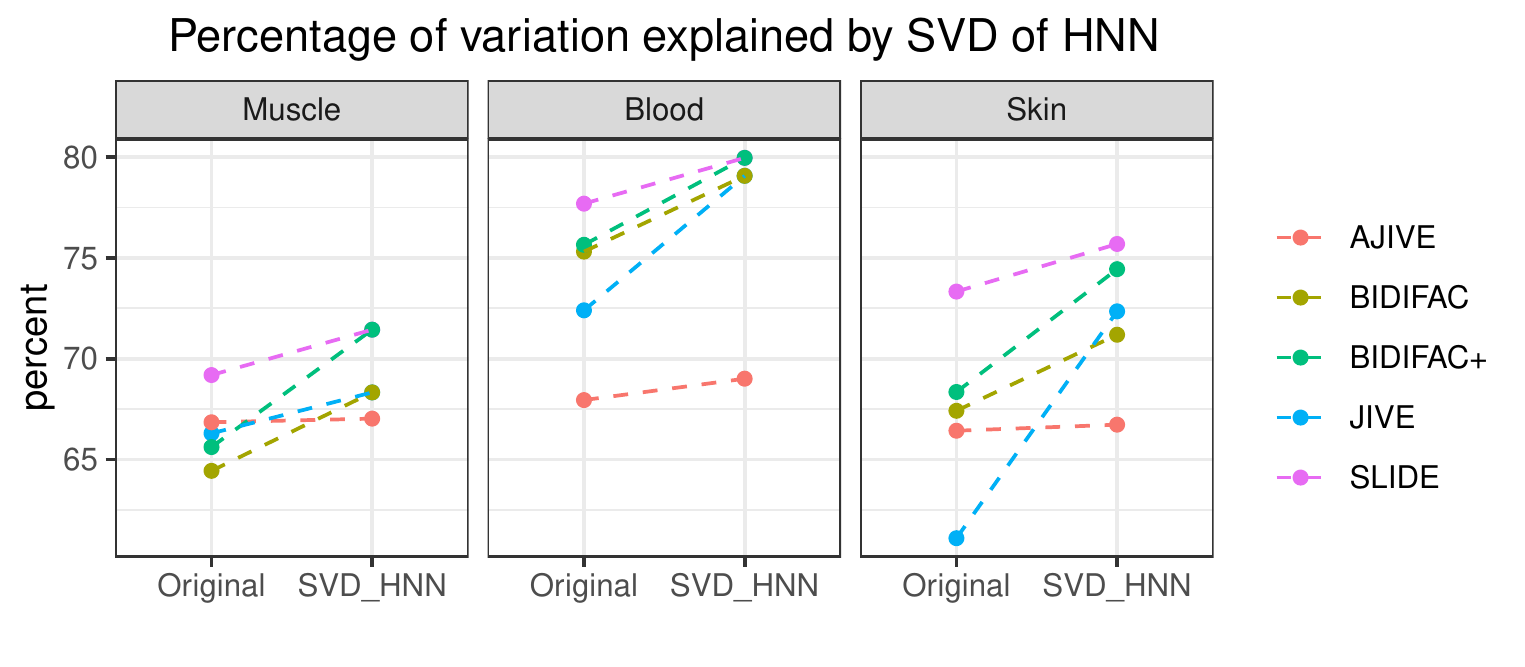} 
    \caption{Each panel shows the percentage of variation explained for each tissue. In the label Original, we show the results of the five competing methods. In the label SVD\_HNN, we present each SVD approximation of the corresponding HNN estimates truncated up to the same rank connected to each method.}\label{fig:svd_GTEx}
\end{figure}

\noindent\textbf{1. Results of each truncated SVD approximation of the HNN estimates} \\
Figure \ref{fig:svd_GTEx} presents the percentage of variation explained by each method and the adjusted HNN estimates based on truncated SVD to match the rank corresponding to each method.

\begin{figure}[H]
    \centering
    \includegraphics[scale = 0.75]{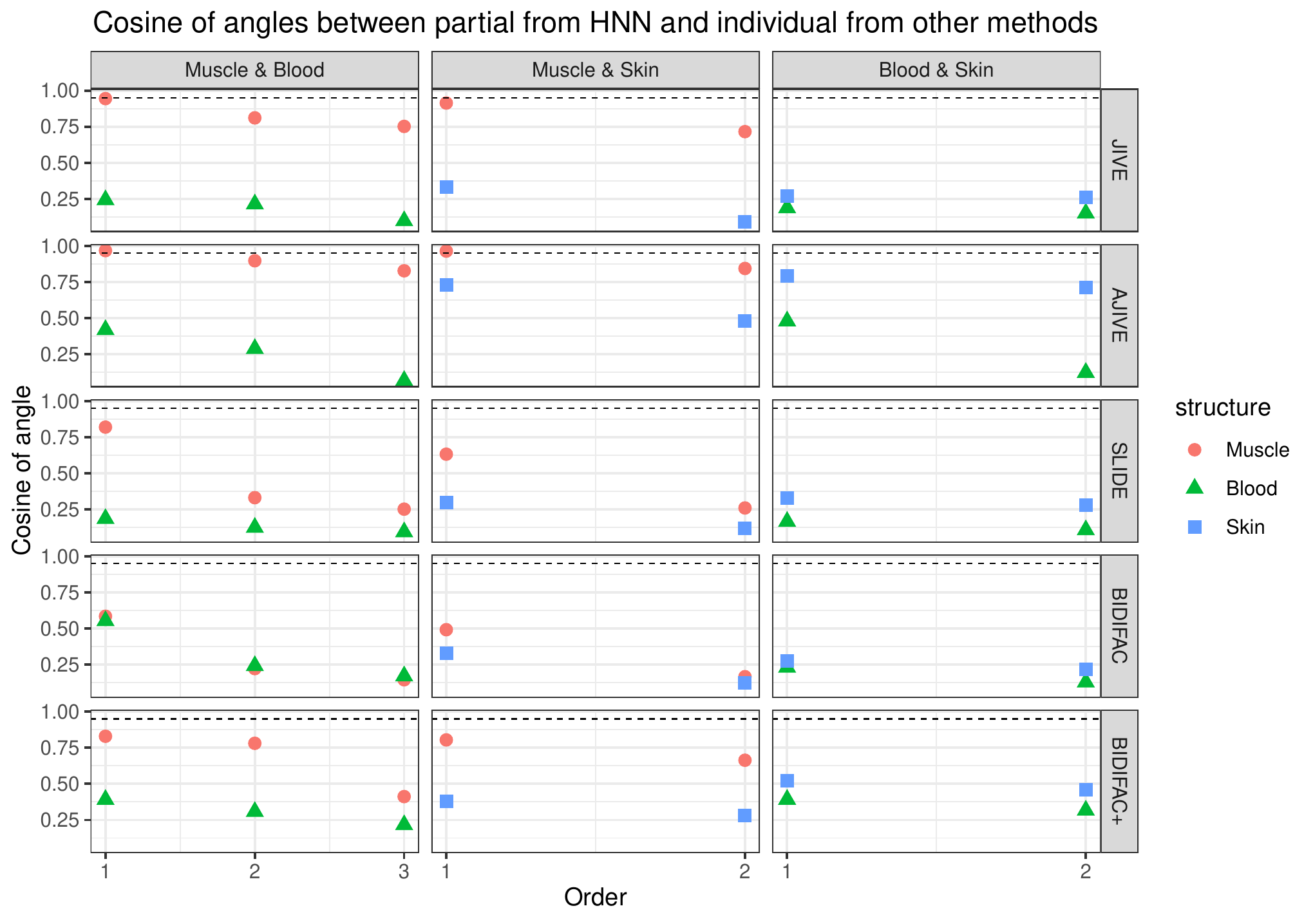} 
    \caption{Cosine of principal angles between each partially-shared structures (1st-3rd columns) from HNN and the individual structures from other methods (1st row: JIVE, 2nd row: AJIVE, 3rd row: SLIDE, 4th row: BIDIFAC, 5th row: BIDIFAC+). For example, the panel labeled as ``Muscle \& Blood” in the 1st row shows the cosine of the principal angles between the partially-shared structures of ``Muscle \& Blood” by HNN and the individual structure of muscle (red circle), the individual structure of blood (green triangle) from JIVE. The dotted horizontal line indicates when the cosine is 0.95.
    }\label{fig:angles2}
\end{figure}

\noindent\textbf{2. Results of principal angles} \\
To see whether the partially-shared signal structures identified by HNN are present in signals estimated by other methods, we compute cosines of principal angles between the partially-shared structures from HNN and the corresponding individual structures from other methods in Figure \ref{fig:angles2}. A cosine value of one indicates zero principal angle (intersection of corresponding subspaces), whereas a cosine value of zero indicates orthogonal signals. Overall, the partially-shared structure by HNN seems to be distinct because, for most cases, both individual structures from each method do not give small angles with each corresponding partially-shared structures by HNN.

\bibliographystyle{apalike} 
\bibliography{all_references}

\end{document}